\documentclass[pra,twocolumn,preprintnumbers,amsmath,amssymb]{revtex4}

\usepackage{graphicx}
\usepackage{dcolumn}
\usepackage{bm}
\usepackage{color}
\usepackage{epstopdf}

\def\eps{\varepsilon}
\def\bk{{\bm \kappa}}

\def\eps{\varepsilon}
\def\bk{{\bm k}}
\def\bx{{\bm x}}
\def\by{{\bm y}}

\begin{document}



\title{Incoherent localized structures and hidden coherent solitons\\ 
from the gravitational instability of the Schr\"odinger-Poisson equation}

\author{Josselin Garnier$^{1}$, Kilian Baudin$^{2}$, Adrien Fusaro$^{2,3}$, 
Antonio Picozzi$^{2}$}
\affiliation{$^{1}$ CMAP, CNRS, Ecole Polytechnique, Institut Polytechnique de Paris, 91128 Palaiseau Cedex, France}
\affiliation{$^{2}$ Laboratoire Interdisciplinaire Carnot de Bourgogne, CNRS, Universit\'e Bourgogne Franche-Comt\'e, Dijon, France}
\affiliation{$^{3}$ CEA, DAM, DIF, F-91297 Arpajon Cedex, France} 


\begin{abstract}
The long-term behavior of a modulationally unstable conservative nonintegrable system is known to be characterized by the soliton turbulence self-organization process: 
A coherent soliton state is eventually generated in order to allow for the increase of the amount of disorder (entropy) in the system.
We consider this problem in the presence of a long-range interaction in the framework of the Schr\"odinger-Poisson (or Newton-Schr\"odinger) equation accounting for the gravitational interaction.
By increasing the amount of nonlinearity, the system self-organizes into a large-scale incoherent localized structure that contains ``hidden" coherent soliton states: The  solitons can hardly be identified in the usual spatial or spectral domains, while their existence is unveiled in the phase-space representation (spectrogram).
We develop a theoretical approach that provides the coupled description of the coherent soliton component (governed by an effective Schr\"odinger-Poisson equation) and of the incoherent component (governed by a wave turbulence Vlasov-Poisson equation).
The theory shows that the incoherent structure introduces an effective trapping potential that stabilizes the hidden coherent soliton, a mechanism that we verify by direct numerical simulations.
The theory characterizes the properties of the localized incoherent structure, such as its compactly supported spectral shape.
It also clarifies the quantum-to-classical correspondence in the presence of gravitational interactions.
This study is of potential interest for self-gravitating Boson models of fuzzy dark matter.
Although we focus our paper on the Schr\"odinger-Poisson equation, we show that our results are general for long-range wave systems characterized by an algebraic decay of the interacting potential.
This work should stimulate nonlinear optics experiments in highly nonlocal nonlinear (thermal) media that mimic the long-range nature of gravitational interactions.
\end{abstract}


\maketitle

\section{Introduction}

The statistical description of nonequilibrium behavior of random dispersive waves is well developed in the {\it weak nonlinear limit}, for which the wave turbulence theory provides a powerful tool to interpret observations arising in contexts as different as ocean waves, quantum fluids, plasmas, and nonlinear optics \cite{zakharov92,Newell01,ZDP04,nazarenko11,newell_rumpf,shrira_nazarenko13,
Turitsyn_rev,PR14}. 
However, such an approach breaks down for {\it strong nonlinearities}, when the turbulent flow can be heavily affected by nonlinear localized excitations, such as the formation of coherent soliton states, as well as condensates, rogue or shock waves \cite{Newell01,ZDP04,nazarenko11,newell_rumpf,brachet11,bortolozzo09,Residori12,Newell13,
shrira_nazarenko13,Turitsyn_rev,TuritsynNP13,TuritsynNC15,PRL11,NC15,PD16,2DturbBEC,PR14,
OnoratoPRE13}.
The understanding of this strong turbulence regime still constitutes a challenging issue of modern physics.

The important idea commonly admitted for nonintegrable systems evolving in the strong nonlinear regime is that the formation of a coherent soliton state plays the role of a ``statistical attractor" for the conservative Hamiltonian system:
It is thermodynamically advantageous to generate a soliton, because this allows the system to increase the amount of disorder (`entropy') in the form of small scale fluctuations \cite{zakharov88,rumpf_newell_01,jordan_josserand00,rumpf_newell_03,Residori12}.
Indeed, the soliton solution realizes the minimum of the energy (Hamiltonian), so that by generating a soliton the system can store the maximum amount of `disorder' by generating rapid small scale fluctuations (kinetic energy contribution).
This scenario is clearly visible through simulations of the focusing nonlinear Schr\"odinger equation.
Starting from a homogeneous state, the wave experiences a (Benjamin-Feir) modulational instability followed by the formation of multiple solitons, which eventually merge into a single large scale coherent soliton that remains immersed in a sea of incoherent small scale fluctuations \cite{zakharov88,rumpf_newell_01,jordan_josserand00,rumpf_newell_03,Residori12}.

The above physical picture becomes more complicated when the system exhibits long-range interactions, that is when the nonlocal nonlinear potential decays algebraically in space.
Indeed, in this case the thermalization process is known to slow down in a dramatic way and a detailed understanding of the dynamics of thermalization is still an open problem, in relation with peculiar features such as violent relaxation, ergodicity breaking, negative specific heat, or inequivalence of thermodynamic ensembles \cite{RuffoPR,ruffo_book}.
In this respect, the Schr\"odinger-Poisson equation (SPE) (or Newton-Schr\"odinger equation) constitutes a natural framework to investigate a wave system featured by long-range interactions.
The SPE was originally proposed to investigate the quantum wave function collapse in the presence of a Newtonian gravitational potential \cite{diosi84,penrose96}. 
It is important to note that the SPE can be derived by considering the non-relativistic limit of the self-gravitating Klein-Gordon equation \cite{ruffini69,giulini12}, while 
soliton solutions of the SPE \cite{chavanis_calmet} have been used in the past to introduce the concept of Bose stars \cite{jetzer92,lee96}. 
In this approach, the formation of a Bose star soliton can be described as a process of gravitational condensation \cite{levkov18,niemeyer20b}. 
The SPE modelling self-gravitating Bose gases in the Newtonian limit also finds an important application in cosmology to describe ``fuzzy dark matter", i.e., dark matter comprised of ultralight scalar bosons whose de Broglie wavelengths are of the order of galaxy scales.
Such a quantum mechanical formulation of dark matter would solve the `cold dark matter crisis', i.e. the formation of a cusp in the classical description of cold dark matter \cite{hu00,peebles03,overduin04,springel05,chavanis11,chavanis_delfini11,suarez14,weinberg15,
witten17,bullock17,niemeyer20}.
This approach is corroborated by several recent three-dimensional (3d) numerical simulations of the SPE performed in the cosmological setting.
They reveal a self-organization process featured by the formation of a large scale soliton core, which is surrounded by an incoherent structure (IS) that is consistent with the classical description 
\cite{schive14a,schive14b,niemeyer16,mocz17,mocz18,levkov18,niemeyer18,niemeyer19,bar18,
mocz19,mocz19b,chavanis19}.
In this perspective, the balance between the gravitational nonlinear potential and the linear dispersion of the wave function (originating from the Heisenberg uncertainty principle) leads to the formation of a soliton core that solves the cusp problem of the classical description of cold dark matter.
We note that, aside from ``fuzzy dark matter", this Bosonic model of dark matter has received different names in the literature, such as wave-dark matter, or ultralight (axion-like) dark matter, scalar field-dark matter or Bose-Einstein condensate-dark matter.

In this article we consider the long-term evolution of the gravitational (modulational) instability of the SPE system.
In a recent work we have shown that, by increasing the amount of nonlinearity, a homogeneous initial condition exhibits the expected gravitational instability and then, after a long transient, the field self-organizes into a localized IS (quasi-stationary state) that contains `hidden' coherent soliton states \cite{PRL21}.
The coherent soliton is `hidden' in the sense that its amplitude is of the same order as the fluctuations of the surrounding IS, while its radius is much larger than the correlation length of the fluctuations of the IS, but much smaller than the radius of the IS.
As a consequence, the coherent soliton state is not easily identified in the usual spatial or spectral domains, but  its existence can be clearly unveiled in the phase-space representation.
Our aim in this article is to pursue the work initiated in Ref.\cite{PRL21} through the analysis of the properties of the IS and the underlying `hidden' coherent solitons. 
Our theoretical approach provides a coupled description of the coherent component (the soliton governed by an effective SPE) and of the IS (the turbulent halo governed by a kinetic equation formally analogous to a wave turbulence (WT) Vlasov-Poisson equation).
The theory shows that the IS introduces an effective trapping potential that stabilizes the hidden coherent soliton, a mechanism that we verify by direct numerical simulations of the coupled system of SPE and WT Vlasov-Poisson equation (VPE).
In addition, the SPE simulations of the random field reveal that the generated IS is characterized by a compactly supported spectral shape, a property that is found in good agreement with a family of stationary analytical solutions of the WT-VPE.
Our theory also clarifies the quantum-to-classical (or SPE to VPE) correspondence in the limit $\hbar/m \to 0$, where the `hidden' solitons appear as the latest residual quantum correction preceding the purely classical limit described by the VPE.
While we develop the theory in the general case of $d-$spatial dimensions, we mainly focus our numerical study in one spatial dimension because for $d=1$ the amount of nonlinearity and the separation of spatial scales (see Eq.(\ref{eq:sep_scales}) below) can be increased in a significant way as compared to 3d simulations -- however note that 2d SPE simulations are also considered in Fig.~\ref{fig:IS_CS_2d}.
Furthermore, although we consider specifically the SPE system, we show that our results are general for long-range wave systems characterized by an algebraic decay of the nonlocal nonlinear potential.

Our work should stimulate nonlinear optics experiments.
Indeed, there is a growing interest in developing analogue gravity phenomena in optical laboratory experiments, i.e., the study of gravitational effects using artificial systems that recreate some specific aspects of the full gravitational system \cite{segev15,faccio_bose_star,faccio_lnp,marino19,nazarenko_nse,navarrete17,paredes20,marino21}.
Our theoretical and numerical predictions should be observed and studied in nonlinear optics experiments involving highly nonlocal nonlinearities \cite{Segev_rev,kivshar_agrawal03,PR14,krolikowski04,NLCliqcryst,conti04,rotschild06,
ghofraniha07,cohen06,rotschild08,vapor,peccianti04,rotschild05,vocke16,marcucci19}.

\section{Schr\"odinger-Poisson equation}

\subsection{Model in $d-$spatial dimensions}

This work is originally motivated by the study of the SPE, which is usually written in the following form 
\begin{eqnarray}
i \hbar \partial_t \psi  &=& - \frac{\hbar^2}{2m}\nabla^2 \psi + m V \psi, 
\label{eq:nse}\\
\nabla^2 V &=& 4 \pi G |\psi|^2.
\label{eq:poiss}
\end{eqnarray}
It describes the 3d evolution of a non-relativistic self-gravitating Bose gas with wave-function $\psi(\bx,t)$ and density $\rho(\bx,t)=|\psi|^2(\bx,t)$.
The non-relativistic Bose gas evolves under the influence of its self-induced gravitational potential $V(\bx,t)$ satisfying the Poisson (or Newton) Eq.(\ref{eq:poiss}), where $m$ is the mass of the Bosons and $G$ the Newton gravitational constant.

In the following we consider a general form of the SPE in spatial dimension $d$:
\begin{eqnarray}
\label{eq:nse_0}
i \partial_{{t}} {\psi} &=& - \frac{\alpha}{2} \nabla^2 {\psi} + {V} {\psi},\\
\label{eq:nse_1}
\nabla^2 {V} &=&  \gamma  \eta_d |\psi|^2,
\end{eqnarray}
with the dispersion coefficient $\alpha > 0$, the nonlinear coefficient $\gamma >0$, and with $\eta_1= 2$, $\eta_2=2\pi$, $\eta_3=4\pi$.
Accordingly, we have 
\begin{equation}
V = -\gamma U_d * |\psi|^2 =  - \gamma \int  U_d({\bx}-{\by}) |{\psi}(\by)|^2 d\by ,
\label{eq:nse_3}
\end{equation}
where the integration is carried out in $\mathbb{R}^d$ with
\begin{equation}
\hspace*{-0.05in}
U_1(x) = -|x|,   \quad U_2(\bx) = -\log(|\bx|), \quad
U_3(\bx)= \frac{1}{|\bx|} .
\label{eq:U_d}
\end{equation}
The SPE conserves the total `mass' 
\begin{eqnarray}
{\cal M}=\int |\psi|^2 (\bx,t) d\bx.
\label{eq:M}
\end{eqnarray}
Note in this respect that the addition of a constant $c_d$ to $U_d$ would only multiply $\psi$ 
by $\exp( i  \gamma c_d {\cal M} t)$, which does not modify the dynamics.
Another important quantity conserved by the SPE is the Hamiltonian ${\cal H}={\cal H}_{l}+{\cal H}_{nl}$, with the linear contribution
\begin{eqnarray}
{\cal H}_{l}(t)=\frac{\alpha}{2} \int |\nabla \psi|^2(\bx,t) d\bx,
\label{eq:H_lin}
\end{eqnarray}
and the nonlinear contribution 
\begin{eqnarray}
{\cal H}_{nl}(t)=\frac{1}{2}\int {V}(\bx,t) |\psi|^2(\bx,t) d\bx .
\label{eq:H_nl}
\end{eqnarray}
It is also worthnoting that the $d-$dimensional SPE (\ref{eq:nse_0}-\ref{eq:nse_3}) is invariant (up to a global phase factor) under the scaling
\begin{eqnarray}
\{ t, \bx, U_d, \psi \} \to 
\{ \mu^{-2}t, \mu^{-1}\bx, \mu^{d-2}U_d, \mu^{2}\psi \}.
\label{eq:inv_scal}
\end{eqnarray}

\subsection{Normalized healing length}
\label{sec:healing_length}

Let us denote by $\bar{\rho}$  the typical amplitude of $|\psi|^2$ and by $\ell$ its typical size, i.e. its typical radius of variation.
Accordingly, the typical amplitude of the gravitational potential $V$ is $\gamma \bar{\rho} \ell^2$ and its typical radius of variation is $\ell$.
Then the characteristic time scale of nonlinear evolution of the system is given by $\tau_{nl}=1/(\gamma \bar{\rho}\ell^2)$.
On the other hand, the characteristic time scale associated to linear dispersion effects is 
$\tau_l = \lambda_c^2/(\alpha/2)$, when the correlation length of the field $\psi$ is $\lambda_c$.
We can then define an analogue of the healing length $\xi$ representing the 
correlation length or spatial scale such that linear and nonlinear effects are of the same order:
\begin{eqnarray}
\xi =  \frac{1}{\ell}\Big( \frac{\alpha}{2\gamma \bar{\rho}}\Big)^{1/2}.
\label{eq:healing_length}
\end{eqnarray}
We remark that $\tau_l/\tau_{nl} = \lambda_c^2/\xi^2$.
Then in the weakly nonlinear regime $\lambda_c \ll \xi$ (or $\tau_l \ll \tau_{nl}$), also called kinetic regime, one can use the WT kinetic theory recently developed for the SPE in Refs.\cite{nazarenko_nse} (see also \cite{levkov18}).
As will be discussed later in detail (see Eq.(\ref{eq:sep_scales})), we will not consider the weakly nonlinear regime in this article.

It is interesting to notice that the SPE does not exhibit free parameters.
To see this, let us consider the following change of variable in Eqs.(\ref{eq:nse_0}-\ref{eq:nse_3}):
$\bx = \Lambda \tilde{\bx}$, $t = \tau \tilde{t}$ and $\psi(t,\bx)=  \sqrt{{\bar \rho}} \tilde{\psi}( t/\tau, \bx/\Lambda)$, with
\begin{equation}
\label{def:tau}
\Lambda = \big( \frac{\alpha}{2\gamma {\bar \rho}} \Big)^{1/4} \mbox{ and }
\tau= \frac{2\Lambda^2}{\alpha}.
\end{equation}
Note that $\Lambda$ denotes the Jeans length, namely the cut-off spatial length below which a homogeneous wave is modulationally stable \cite{comment1}.
The normalized field $\tilde{\psi}$ then satisfies the dimensionless and normalized SPE
\begin{eqnarray}
i \partial_{\tilde{t}}\tilde{\psi}  =- \nabla^2 \tilde{\psi} - \tilde{\psi}  \int  U_d(\tilde{\bx}-\tilde{\by}) |\tilde{\psi}(\tilde\by)|^2 d\tilde\by 
\end{eqnarray}
in $\mathbb{R}^d$. 
The normalized healing length $\tilde{\xi} = \xi / \Lambda$ then reads 
\begin{eqnarray}
\tilde{\xi} = \frac{\Lambda}{\ell} .
\label{eq:healing_length_norm}
\end{eqnarray}
Note that this dimensionless parameter is invariant under the scaling (\ref{eq:inv_scal}) of the $d-$dimensional SPE (\ref{eq:nse_0}-\ref{eq:nse_3}).

\section{Numerical simulations}
\label{sec:simul_0}

For the sake of clarity, we start this section by recalling the properties of the regime of `hidden' solitons in relation with the quantum to classical transition, as recently discussed in Ref.\cite{PRL21}.
Next, through the analysis of SPE numerical simulations, we discuss the properties that characterize the large scale IS that surrounds and supports the hidden solitons.

\subsection{Quantum to classical correspondence}

The parameter $\tilde{\xi}$ defined by Eq.(\ref{eq:healing_length_norm}) is directly related to a parameter $\Xi$ that was shown to control the quantum to classical transition (i.e. SPE to Vlasov-Poisson correspondence) in the limit $\hbar/m \to 0$: 
By Eq.(42) in \cite{mocz18} and by Eq.(\ref{eq:norm_hl_nse}) below, we have $\Xi \sim \tilde{\xi}^4$.
Actually, the parameter $\Xi$ was introduced on the basis of scale-free invariants and dimensional arguments for the three-dimensional case $d=3$ \cite{niemeyer16,mocz17}.
Accordingly, for moderate values of ${\tilde \xi} \sim 1$, the linear dispersion term in the SPE plays a fundamental role and the system exhibits a coherent dynamics that is essentially dominated by soliton-like structures. 
This aspect has been studied through intensive numerical simulations in the cosmological setting \cite{niemeyer20,schive14a,schive14b,niemeyer16,mocz17,mocz18,niemeyer18,niemeyer19,bar18,
mocz19,mocz19b,schive20}, which revealed the formation of an IS that is dominated in its center by a large amplitude coherent soliton peak $\rho_S$, typically larger than the average density of the surrounding IS \cite{bar18}.
In addition, the soliton radius $R_S$ is typically of the order of the correlation radius of the fluctuations of the IS, $R_S \sim \lambda_c$ \cite{schive14a,schive14b,niemeyer16,mocz17,mocz18,mocz19,niemeyer20b}.
By decreasing the parameter ${\tilde \xi} \ll 1$ (or equivalently $\Xi \ll 1$), soliton states are expected to gradually disappear.
The system enters a collective regime, featured by the generation of a large scale IS \cite{mocz18}.

In the recent work \cite{PRL21}, we showed that such an IS can still contain `hidden' coherent soliton states.
More precisely, we showed that in the regime ${\tilde \xi} \ll 1$ (or $\Xi \ll 1$), an IS with amplitude $\bar{\rho}$, typical radius $\ell$ and characterized by a correlation length $\lambda_c \sim \xi$ can stabilize a coherent soliton of amplitude $\bar{\rho}$ and typical radius $R_S$ verifying 
\begin{eqnarray}
 \lambda_c \sim \xi \ll R_S \ll \ell.
\label{eq:sep_scales}
\end{eqnarray}
Actually, $R_S \sim \Lambda=\sqrt{\xi \ell}$ is the harmonic  average of  $\ell$ and $\xi$.
A more precise result will be obtained in this article by combining the stationary solution of the WT-VPE with the soliton mass-radius relation, see below Eqs.(\ref{eq:corresp_3d_0}-\ref{eq:corresp_3d}).

\begin{center}
\begin{figure}
\includegraphics[width=1\columnwidth]{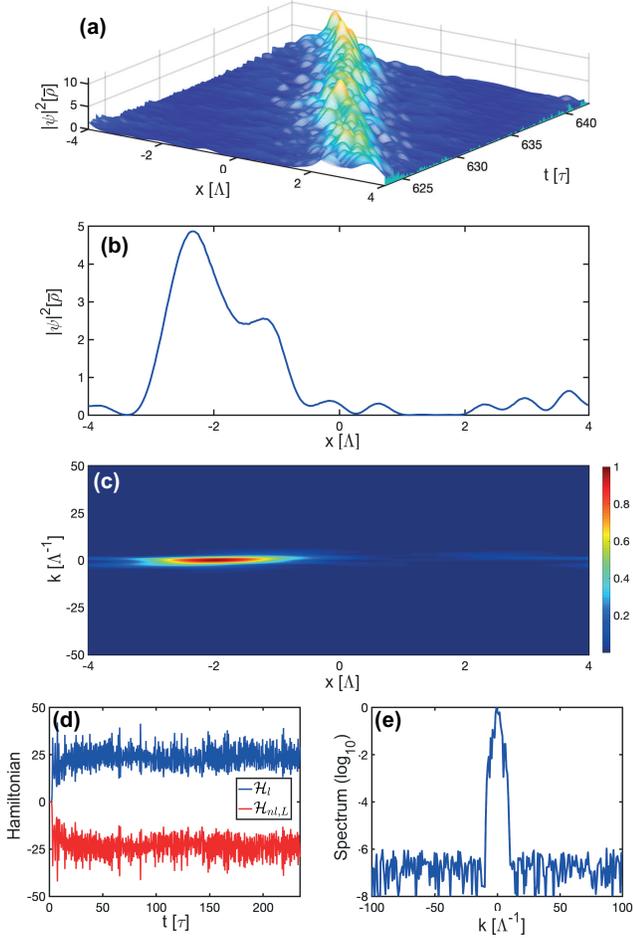}
\caption{
\baselineskip 10pt
{\bf Coherent soliton-like regime:}  
Numerical simulation of the SPE (\ref{eq:nse_0}-\ref{eq:nse_3}) 
in the one-dimensional spatial domain $[-L/2,L/2]$ with periodic boundary conditions starting from a homogeneous initial condition with density $\bar{\rho}$. 
Here $L=8\Lambda$
so that  ${\tilde \xi}\simeq 0.25$.
The initial wave develops the gravitational (modulational) instability and after a transient the system exhibits a coherent soliton-like dynamics with random wave fluctuations.
Spatio-temporal evolution of the density $|\psi|^2(x,t)$ (a), spatial profile at $t=64 \tau$ (b) and corresponding phase-space spectrogram (Husimi transform) (c).
(d) Evolutions of the linear ${\cal H}_l$ and nonlinear energies ${\cal H}_{nl,L}$.
(e) Spectrum of the wave $|{\tilde \psi}|^2(k)$ at $t=64 \tau$, where ${\tilde \psi}(k)$ is the Fourier transform of $\psi(x)$.
See Eq.(\ref{def:tau}) for the definition of $\Lambda$ and $\tau$, see Eq.(\ref{def:HnlL}) for the definition of ${\cal H}_{nl,L}$.
}
\label{fig:sol_regim_1d} 
\end{figure}
\end{center}

\begin{center}
\begin{figure}
\includegraphics[width=1\columnwidth]{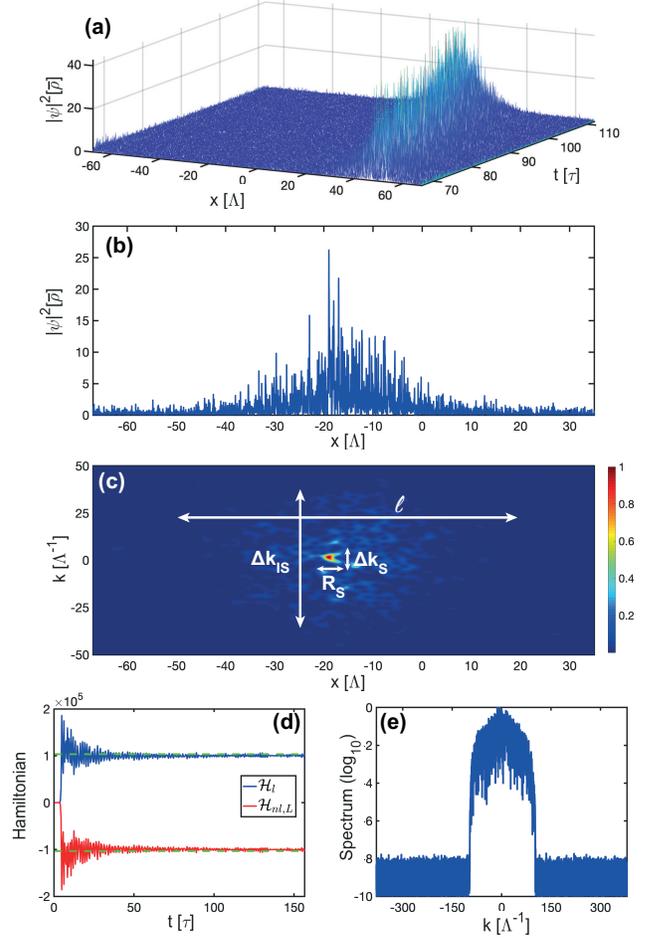}
\caption{
\baselineskip 10pt
{\bf Incoherent structure with hidden coherent soliton:}
Simulation of the SPE (\ref{eq:nse_0}-\ref{eq:nse_3})
in the one-dimensional spatial domain $[-L/2,L/2]$ with periodic boundary conditions starting from a homogeneous initial condition with density $\bar{\rho}$.
Here $L=135\Lambda$ so that 
${\tilde \xi} \simeq 1.5\times 10^{-2}$.
The wave exhibits the gravitational instability and after a transient it self-organizes into a localized IS that contains hidden coherent solitons.
Spatio-temporal evolution of the density $|\psi|^2(x,t)$ (a), spatial profile at $t=143\tau$ (b) and corresponding phase-space spectrogram (Husimi transform) (c).
The size of the soliton ($R_S$) is much smaller than the size of the IS, $R_S \ll \ell \sim L/2$, while $R_S$ is much larger than the correlation length of the fluctuations of the IS, $R_S \simeq 2\pi/\Delta k_{S} \gg \lambda_c \simeq 2\pi/\Delta k_{IS}$, see Eq.(\ref{eq:sep_scales_spectro}).
(d) Evolutions of the linear ${\cal H}_l$ and nonlinear ${\cal H}_{nl,L}$ energies showing the relaxation toward a quasi-stationary state.
The green dashed lines denote the theoretical values of ${\cal H}_l$ and ${\cal H}_{nl,L}$ predicted by the stationary solution of the WT-VPE, see Eq.(\ref{eq:Hlnltheo}).
(e) Wave spectrum $|{\tilde \psi}|^2(k)$ at $t=143\tau$ (${\tilde \psi}(k)$ being the Fourier transform of $\psi(x)$) -- note that the spectrum exhibits a compact support.
}
\label{fig:inc_sol_regim_1d} 
\end{figure}
\end{center}

\subsection{SPE simulations}

As discussed in the introduction section, we address more specifically the problem of a gravitational unstable homogeneous initial condition of the wave amplitude.
Note that the homogeneous initial condition is relevant to standard cosmology where the formation of large scale structures such as galaxies and galaxy clusters proceeds via gravitational instabilities from small primordial density fluctuations \cite{mocz19,malomed85,binney}.
We consider the gravitational instability through numerical simulations of the SPE in a finite box of size $L^d$ with periodic boundary conditions and uniform initial density, see Appendix~\ref{app:A}.  
As discussed in Ref.\cite{PRL21}, the parameter ${\tilde \xi}$ in (\ref{eq:healing_length_norm}) (computed with $\ell \simeq L/2$) does not depend on the spatial dimension $d$.
Since numerical simulations in three spatial dimensions ($d=3$) are extremely time consuming, it proves convenient to consider the regime ${\tilde \xi} \ll 1$ in one spatial dimension $d=1$.
In other words, for $d=1$ it is possible to carry out numerical simulations with a large separation of the three scales $\lambda_c \ll R_S \ll L$ (see Eq.(\ref{eq:sep_scales})), a feature which does not appear possible for $d=3$ with the present computational resources.
Indeed, we consider below values of $\Xi \simeq 5 \times 10^{-8}$ for $d=1$, while typically $\Xi > 10^{-4}$ for $d=3$ in Refs.\cite{mocz17,mocz18}.
Note that SPE simulations in two spatial dimensions ($d=2$) are also reported in the Appendix~\ref{app:G} for moderate values of ${\tilde \xi}$ that are accessible in our simulations, which evidence the formation of a soliton that is only partially hidden within the IS.

The initial condition is a homogeneous wave $\psi(x,t=0)=\sqrt{{\bar \rho}}$ with a superimposed small noise to initiate the gravitational (modulational) instability process.
As broadly discussed in the literature 
\cite{schive14a,schive14b,niemeyer16,mocz17,mocz18,niemeyer19,niemeyer18,mocz19}, the gravitational instability undergoes a collapse, which is eventually regularized by the formation of a virialized incoherent halo.

We report in Figs.~\ref{fig:sol_regim_1d}-\ref{fig:inc_sol_regim_1d} the results of SPE simulations for two different values of the normalized healing length ${\tilde \xi} \simeq 0.25$ and ${\tilde \xi} \simeq 1.5\times 10^{-2}$ (that are computed with $\ell=L/2$).
The system exhibits two radically different regimes.
For relatively `large' values of 
${\tilde \xi} \sim 1$, the system exhibits a coherent phase-sensitive regime that is essentially dominated by soliton-like  structures, see Fig.~\ref{fig:sol_regim_1d}.
In this regime ${\tilde \xi} \sim 1$, the typical soliton width is comparable to the correlation length of the IS, which is itself of the order of the size of the IS.

The regime of interest in this paper arises by decreasing the parameter ${\tilde \xi}$. 
This regime is characterized by the emergence of a large scale IS that is localized in space, as illustrated in Fig.~\ref{fig:inc_sol_regim_1d}.
Note that Fig.~\ref{fig:inc_sol_regim_1d}(a) shows the long-term evolution once the IS has reached a quasi-staionary state.
This is illustrated by the relaxation process evidenced in the evolutions of the linear and nonlinear Hamiltonian in panel Fig.~\ref{fig:inc_sol_regim_1d}(d).
A key difference that distinguishes the IS in Fig.~\ref{fig:inc_sol_regim_1d} and in Fig.~\ref{fig:sol_regim_1d} is that for small values of ${\tilde \xi}$ the IS does not exhibit apparent coherent soliton structures. 
This observation appears consistent with the fact that the SPE (\ref{eq:nse}-\ref{eq:poiss}) should recover the classical Vlasov-Poisson equation in the limit $\hbar/m \to 0$ \cite{mocz18}: Considering the SPE (\ref{eq:nse}-\ref{eq:poiss}) where $\alpha=\hbar/m$ and $\gamma= m G/\hbar$, the normalized healing length (\ref{eq:healing_length_norm}) 
reads 
\begin{eqnarray}
{\tilde \xi}= \frac{(\hbar/m)^{1/2}}{\ell (2 {\bar \rho} G)^{1/4}}.
\label{eq:norm_hl_nse}
\end{eqnarray}
${\tilde \xi}$ is then directly related to a parameter that controls the SPE to Vlasov-Poisson correspondence, namely $\Xi \sim  {\tilde \xi}^4$ \cite[Eq.(42)]{mocz18}.
Accordingly, the limit ${\tilde \xi} \to 0$ corresponds to the classical Vlasov-Poisson equation, which is inherently unable to describe coherent soliton structures.
However, as discussed in Ref.\cite{PRL21}, the simulations reveal that the IS in Fig.~\ref{fig:inc_sol_regim_1d} is not purely incoherent, but contains hidden coherent soliton structures.
Such a coherent soliton is unveiled by a phase-space analysis of the field $\psi(x)$ provided by the Husimi representation, which is in substance a smoothed version of the Wigner transform, see  Ref.\cite{mocz18} or the Supplementary Material in Ref.\cite{PRL21}.
Indeed, the Wigner  transform in $d$-spatial dimension 
$W(\bk,\bx,t)= \int \psi(\bx+\by/2,t)  \psi^*(\bx-\by/2,t)\exp(-i {\bm k} \cdot\by) \, d\by$,
is not statistically stable, in the sense that its standard deviation is larger than its statistical average.
We consider the smoothed Wigner transform or Husimi function
\begin{eqnarray}
W_\sigma (\bk,\bx,t)= \frac{1}{\pi^d} \iint W(\bk',\bx',t)  \quad \quad \quad \quad \quad \quad \nonumber \\
\quad \quad \quad \times \exp\Big( - \frac{|\bx-\bx'|^2}{\sigma^2} - \sigma^2 |\bk-\bk'|^2\Big) 
d\bk' d\bx',
\label{eq:husimi}
\end{eqnarray}
which is statistically stable for all $\sigma>0$.
A small (large) $\sigma$ means that the smoothing is smaller (larger) in $\bx$ than in $\bk$  by (\ref{eq:husimi}).
The Husimi transform can also be written as a `local' Fourier transform
\begin{eqnarray*}
W_\sigma (\bk,\bx,t)= \frac{1}{\pi^{d/2}}
\Big| \int \psi(\by) {\cal G}_\sigma(\bx-\by) \exp( - i \bk \cdot \by) d\by  \Big|^2   ,
\label{eq:husimi2}
\end{eqnarray*}
where ${\cal G}_\sigma(\bx)=\sigma^{-d/2} \exp[-|\bx|^2/(2\sigma^2) ]$.
It is important to note that, in optics, the Husimi transform is provided by the experimental measurement of the spatial spectrogram \cite{waller12}.
Figure~\ref{fig:inc_sol_regim_1d}(c), shows the 1d Husimi transform of $\psi(x)$ at a particular time.
The coherent soliton is characterized in phase-space by a high intensity spot immersed in a sea of small scale fluctuations. 
Notice that the soliton has a spectral width $\Delta k_{S}$ much smaller than the spectral width of the IS, $\Delta k_{IS} \gg \Delta k_{S}$, which means that the width of the soliton, $R_S \simeq 2\pi/\Delta k_{S}$, is much larger than the correlation length $\lambda_{c} \simeq 2\pi/\Delta k_{IS}$ of the IS.
In addition, the radius of the soliton $R_S$ is much smaller than that of the IS $\ell$.
Then we observe in Fig.~\ref{fig:inc_sol_regim_1d}(c) the separation of spatial scales in (\ref{eq:sep_scales}), namely
\begin{eqnarray}
\lambda_{c} \simeq 2\pi/\Delta k_{IS} \ll R_S \simeq 2\pi/\Delta k_{S}  \ll \ell \sim L/2.
\label{eq:sep_scales_spectro}
\end{eqnarray}
This hidden soliton regime is in marked contrast with the coherent regime discussed above for ${\tilde \xi} \sim 1$.
Indeed, as revealed by the corresponding phase-space spectrogram reported in Fig.~\ref{fig:sol_regim_1d}(c), in the coherent soliton regime ${\tilde \xi} \sim 1$ there is no scale separation between the radius of the soliton and the radius of the whole localized structure.
We refer the reader to Ref.\cite{PRL21} for a detailed discussion of the dynamics of the hidden coherent solitons embedded in the IS in the regime ${\tilde \xi} \ll 1$.

\subsection{The incoherent structure beyond the weakly nonlinear regime}

The SPE simulations reveal a remarkable property of the IS, namely that it is characterized by a spectrum that exhibits a compact support, see Fig.~\ref{fig:inc_sol_regim_1d}(e).
This indicates the existence of a frequency cut-off $k_c$, whose origin can be explained from self-consistent stationary properties for the IS, see below section~\ref{sec:stat_sol_vlasov_1d}.
Such a compactly supported spectral shape is in  contrast with an algebraic {\it power-law spectrum $n_k \sim 1/k^\nu$}, which is a stationary solution of the (collisional) WT kinetic equation \cite{nazarenko_nse} -- see in particular the case $\nu=2$ for the thermal equilibrium solution (Rayleigh-Jeans distribution).
Accordingly, the IS discussed here corresponds to a regime that is very far from thermal equilibrium.
Actually, the IS is very robust and lasts for large interaction times, as illustrated by the evolutions of the linear energy (${\cal H}_{lin}$) and the nonlinear energy (${\cal H}_{nl,L}$) reported in Fig.~\ref{fig:inc_sol_regim_1d}(d).
Hence, we refer this collective incoherent state to a quasi-equilibrium state, in analogy with the {\it quasi-stationary states} investigated in long-range interacting systems  \cite{RuffoPR,ruffo_book}.

It is also important to note that the IS discussed in this work evolves in the {\it strong nonlinear interaction regime}, i.e., it is characterized by a correlation length of the same order as the healing length, $\lambda_c \sim \xi$ (or $\tau_l \sim \tau_{nl}$).
In other words, the highest frequency components of the IS are of the order $k_c \sim 1/\xi$ and thus do not evolve in the weakly nonlinear (kinetic) regime. 
This was anticipated above through the separation of spatial scales (\ref{eq:sep_scales}), and it will be confirmed by the theory in Eqs.(\ref{eq:k_c_1d}-\ref{eq:corr_length_IS}) and the simulations, see Fig.~\ref{fig:NLSvsVlasov}.
In addition, the IS is characterized by spatial fluctuations that are not homogeneous in space.
This means that the description of the IS is not captured by a weakly nonlinear kinetic WT formalism \cite{nazarenko_nse}.
In the next section we show that the appropriate description of the IS is provided by the long-range WT VPE.

\section{Effective SPE and WT-VPE formalisms}
\label{sec:coupled_NLS_vlasov}

\subsection{Coupled SPE and WT-VPE equations}

In this section we provide a formalism that describes the coupled coherent-incoherent dynamics governing the evolution of the IS and the underlying soliton.
The IS will be described theoretically within the general framework of the WT formalism \cite{PR14,PRL11}. 
The WT theory has been shown to provide a natural asymptotic closure of the hierarchy of moment equations for a system of weakly nonlinear dispersive waves \cite{zakharov92,Newell01,ZDP04,nazarenko11,newell_rumpf}.   
Here, the IS is characterized by fluctuations that are not homogeneous in space, so that, at leading order the dynamics is dominated by a long-range version of the WT-VPE \cite{PRL11,PR14,NC15,PD16}.
Note that the WT-VPE differs from the traditional Vlasov equation describing random waves (e.g., incoherent modulational instabilities, or incoherent solitons) in optics \cite{PR14,Segev_rev}, hydrodynamics \cite{Newell13,OnoratoPRE13} or plasmas \cite{lvov77,ZakharovPR85,nazarenko92}, while its structure is analogous to that describing systems of particles with long-range, e.g. gravitational, interactions \cite{RuffoPR}.
In addition, at variance with conventional weak-turbulence approaches \cite{zakharov92,ZDP04}, the long-range WT-VPE is valid beyond the weakly nonlinear regime of interaction \cite{PRL11}, as revealed by a recent study of incoherent shock waves that develop in the strong nonlinear regime \cite{NC15}, also see \cite{PRA12,OL14}.
Indeed, thanks to the long-range nature of the interaction, the system exhibits a self-averaging
property of the nonlinear response, $ \int  U_d({\bx}-{\by}) |{\psi}(\by)|^2 d\by \simeq \int  U_d({\bx}-{\by}) \left<|{\psi}(\by)|^2\right> d\by$. 
Substitution of this property into the SPE thus leads to an automatic closure of the hierarchy of the moment equations. 
More specifically, using statistical arguments similar to those in Ref.\cite{garnier03}, one can show that, owing to a highly nonlocal response, the statistics of the incoherent wave turns out to be Gaussian.
We will see in the following that the long-range WT-VPE formalism provides an accurate description of the IS observed in the simulations, which evolve in the strong nonlinear regime $\lambda_c \sim \xi$ (see Eqs.(\ref{eq:k_c_1d}-\ref{eq:corr_length_IS}) and Fig.~\ref{fig:NLSvsVlasov} below).


In order to describe the coupled coherent-incoherent dynamics of the IS and the underlying soliton, here we derive a coupled system of SPE and WT-VPE.
Indeed, the soliton is characterized by a non-vanishing average $\left< \psi \right> \neq 0$, so that at variance with the usual derivation of the WT-Vlasov equation, here we decompose the field into a coherent component $A(\bx,t)$ and an incoherent component 
$\phi(\bx,t)$ of zero mean ($A=\left<A\right> \neq 0, \left< \phi \right> =0$):
\begin{eqnarray}
\psi(\bx,t)=A(\bx,t)+\phi(\bx,t).
\label{eq:decomposepsi}
\end{eqnarray}
Note that this approach is similar to the two-fluid model developed to describe superfluid helium as a mixture of two fluids, the normal and the superfluid components \cite{nazarenko92}.
Following the usual procedure \cite{PR14}, we define the `local' spectrum of the IS as the {\it average} Wigner transform 
$n({\bm k},{\bm x},t)= \int  B(\bx,{\bm \xi},t)  \, \exp(-i {\bm k} \cdot {\bm \xi}) \, d{\bm \xi}$, where the correlation function 
\begin{eqnarray}
B(\bx,{\bm \xi},t)=\left< \phi(\bx+{\bm \xi}/2,t)  \phi^*(\bx-{\bm \xi}/2,t)\right>,
\end{eqnarray}
is defined from an average over the realizations $\left< \cdot \right>$.
It is important to distinguish the averaged quantity $n({\bm k},{\bm x})$ from the {\it non-averaged} Wigner transform $W({\bm k},{\bm x})$ discussed above through the Husimi transform.
Starting from the SPE we obtain 
\begin{align}
& i\partial_t A  = - \frac{\alpha}{2} \nabla^2 A + A V , 
\label{eq:nls_0}
\\
& \partial_t n({\bm k},{\bm x}) +\alpha {\bm k}\cdot \partial_{\bx} n({\bm k},{\bm x}) - \partial_{\bm x} V \cdot \partial_{\bm k} n({\bm k},{\bm x}) = 0 ,
\quad \quad 
\label{eq:vlasov_0}
\end{align}
which are coupled to each other by the {\it averaged} long-range gravitational potential 
\begin{align}
& V({\bm x},t) = - \gamma \int U_d({\bm x}-{\bm y})  \big( |A|^2({\bm y},t) + N({\bm y},t) \big) d{\bm y}, 
\label{eq:V_0} \\
& N({\bm x},t)=\left< |\phi(\bx,t)|^2 \right>=\frac{1}{(2 \pi)^d}  \int n({\bm k},{\bm x},t) d \bk. 
\label{eq:N_0} 
\end{align}
Note that $N({\bm x},t)$ denotes the {\it average} density of the IS, which depends on the spatial variable $\bx$ because the IS exhibits fluctuations that are not homogeneous in space -- it should not be confused with the {\it non-averaged} density $\rho(\bx,t)=|\psi(\bx,t)|^2$ (see e.g. Fig.~\ref{fig:NLSvsVlasov} where $\rho(\bx)$ and $N({\bm x})$ are superposed).

Finally, it is also important to distinguish the WT-VPE (\ref{eq:vlasov_0}) from the collisionless Boltzmann VPE \cite{mocz18,uhlemann14}.
The WT-VPE describes the smooth evolution of the second-order moment $n(\bk,\bx)$ defined from the {\it average} over the realizations $\left< \cdot \right>$ of the random wave. 
This is in contrast with the {\it spiky distribution} solution of the collisionless Boltzmann kinetic equation, see \cite{mocz18,schive14a}.
To avoid confusion, in this paper we denote by WT-VPE the derived kinetic Eq.(\ref{eq:vlasov_0}) governing the smooth evolution of $n(\bk,\bx)$.

\subsection{Multi-scale analysis: Rescaled coupled SPE and WT-VPE}
\label{sec:scaling}

The system of coupled SPE and WT-VPE (\ref{eq:nls_0}-\ref{eq:N_0}) is quite complicated.
However, the regime of hidden solitons identified in the simulations is characterized by the small parameter $\eps \equiv {\tilde \xi}\ll 1$.
Accordingly, further insight can be obtained by using a multi-scale series expansion in the small parameter $\eps$.
We showed in Ref.\cite{PRL21} that the regime of hidden solitons is described by the following scaling:
\begin{eqnarray}
A(\bx,t) &=& A^{(0)}(\bx,t),
\label{eq:A_0} \\
n(\bk,\bx,t) &=& \eps^d n^{(0)}(\eps \bk, \eps \bx, t).
\label{eq:n_0} 
\end{eqnarray}
With this scaling we have $N(\bx,t) = N^{(0)}(\eps \bx,t)$ (with $N^{(0)}(\eps \bx,t)=(2\pi)^{-d}\int n^{(0)}(\eps \bx,\eps \bk,t)d(\eps \bk)$).
This means that the soliton density $|A|^2(\bx)$ is of the same order as the average density $N(\bx)$ of the IS, i.e. the soliton is hidden in the IS. 
However the soliton is unveiled in the phase-space representation $(\bx,\bk)$ as a peak of order one as compared to the background amplitude of $n(\bk,\bx)$ that is of order $\eps^d$, see Fig.~\ref{fig:sol_regim_1d}(e).
In this way, the scaling (\ref{eq:A_0}-\ref{eq:n_0}) in the spatial variable $\bx$ means that the radius of the large scale IS is of order $\sim 1/\eps \gg 1$ compared to the radius of the soliton that is of order one.
On the other hand, the scaling (\ref{eq:A_0}-\ref{eq:n_0}) in the wavevector variable $\bk$ means that the correlation length of the IS $\lambda_{c} \sim 1/\Delta k_{IS}$ is of order $\eps \ll 1$ compared to the radius $R_S$ of the soliton that is of order one. 
In addition, the mass of the IS is $M_{IS}=\eps^{-d} \int N^{(0)}(\eps \bx) d(\eps \bx)$, which is of order  $\sim \eps^{-d}$ compared to the mass of the soliton $M_{S}=\int |A^{(0)}(\bx)|^2 d\bx$, which is of order one.
Accordingly
\begin{eqnarray}
M_{IS} \gg M_{S},
\label{eq:rapp_masses}
\end{eqnarray}
and the IS provides the main contribution to total mass ${\cal M} \simeq M_{IS}$. 
This separation of scales between the soliton and the IS was anticipated through the numerical simulations.
The theory we are going to present will confirm this expectation.

The multi-scale expansion procedure reported in Ref.\cite{PRL21} shows that the coupled SPE and WT-VPE (\ref{eq:nls_0}-\ref{eq:vlasov_0}) take a rescaled form 
\begin{eqnarray}
&& i\partial_t A  = - \frac{\alpha}{2} \nabla^2  A + V_{S} A + \gamma  q_d   N({\bf 0},t) |\bx|^2 A , \quad
\label{eq:nls_A0} \\
&& V_{S}(\bx,t)= - \gamma \int U_d(\bx-\by) |A|^2(\by,t) d\by,
\label{eq:V_0_coh} 
\end{eqnarray}
where the constant $q_d$ depends on the spatial dimension, $q_3=2\pi/3, q_2=\pi/2, q_1=1$.
On the other hand, the WT-VPE (\ref{eq:vlasov_0}) reduces to 
\begin{eqnarray}
&& \partial_t n(\bk,\bx) +\alpha \bk \cdot \partial_{\bx} n(\bk,\bx) 
- \partial_{\bx} V_{IS} \cdot \partial_{\bk} n(\bk,\bx) = 0 ,
\quad \quad
\label{eq:vlasov_N0} \\
&& V_{IS}(\bx,t) = - \gamma \int U_d(\bx-\by)N(\by,t) d\by,
\label{eq:V_0_inc}
\end{eqnarray}
where we recall that $N(\bx,t)=(2\pi)^{-d}\int n(\bk,\bx,t) d\bk$.

This analysis then reveals three important results:

\noindent
(i) The rescaled WT-VPE (\ref{eq:vlasov_N0}) governing the evolution of the IS is not affected by the hidden soliton, since the self-consistent gravitational potential $V_{IS}$ does not depend on the soliton, but solely on the IS described by $N(\bx,t)$.
This result is consistent with the fact that the mass of the soliton is negligible with respect to the mass of the IS, as discussed through (\ref{eq:rapp_masses}).

\noindent
(ii) The hidden soliton experiences its self-gravitational potential $V_{S}$, as well as a trapping parabolic  potential, as revealed by the last term in Eq.(\ref{eq:nls_A0}).
As discussed in detail in Ref.\cite{PRL21}, this parabolic trapping potential induced by the IS  explains the phase-space circular motion of the soliton that are observed in the simulations, as well as the formation of binary solitons that spiral around each other in phase-space.

\noindent
(iii) The mass of the soliton and the mass of the IS are conserved during the evolution, i.e., there is no mass exchange among each other.
This result is valid, strictly speaking, for an `infinite' separation of the spatial scales $\lambda_c \ll R_S$.
However, a weak exchange of mass between the soliton and the IS can be observed in the simulations, which is due to a `finite' separation of the scales $\lambda_c$ and $R_S$.
Then the evolution of the spectrogram $n(\bk,\bx,t)$ of the IS is not, strictly speaking, independent of the soliton dynamics. 
Actually, the IS spectrum feels the presence of the soliton (see Eqs.(\ref{eq:vlasov_0}-\ref{eq:V_0})), which in turn affects the effective potential seen by the soliton.
The soliton then sees a potential that is not purely quadratic but locally modified by its own presence, which introduces small fluctuations in the soliton dynamics and thus a weak exchange of mass with the IS.

Along this way, when the spatial scales $\lambda_c$ and $R_S$ are hardly separated,  the prediction of the existence of stable soliton states provided by the effective SPE (\ref{eq:nls_A0}-\ref{eq:V_0_coh}) \cite{PRL21} is only valid for relatively small interaction times.
Indeed, SPE simulations evidence a degradation of the soliton structure for large interaction times.
Actually, the issue of the long-time stability of the hidden soliton is a difficult problem, which should be analyzed through the inspection of the higher-order terms  neglected in the multi-scale analysis with the small parameter ${\tilde \xi}$.

\section{Properties of the incoherent localized structure}
\label{sec:propertiesIS}

\subsection{Stationary solution for the incoherent structure}
\label{sec:stat_sol_vlasov_1d}

\begin{center}
\begin{figure}
\includegraphics[width=.8\columnwidth]{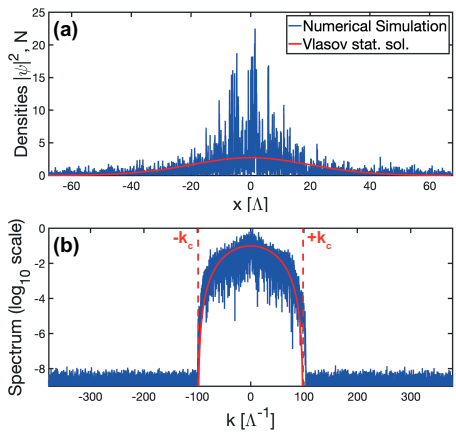}
\caption{
\baselineskip 10pt
{\bf Incoherent localized structure: SPE simulation vs WT-VPE stationary solution.}
Numerical simulation of the SPE (blue line) with ${\tilde \xi} \simeq 1.5\times 10^{-2}$ (as in Figure \ref{fig:inc_sol_regim_1d},  $L=135\Lambda$) showing the density $|\psi|^2(x)$ (a), and the spectrum $|{\tilde \psi}|^2(k)$ in log$_{10}$-scale (b), at $t=188 \tau$.
Starting from a homogeneous initial condition, the wave self-organizes into an incoherent localized structure (a), characterized by a compact support of the spectrum (b).
Comparison with the stationary solution of the  WT-VPE (\ref{eq:stat_sol_vlasov}) (red line), whose parameters ($p=4.76, N_{st}(0)=2.72 {\bar \rho}$) have been obtained from 
Eqs.(\ref{eq:valuep}-\ref{eq:valueNst0}).
Note in particular the good agreement with the compactly supported spectral shape and the corresponding frequency cut-off $k_c \simeq 1.03 {\sqrt{\gamma \bar{\rho}} L}/{\sqrt{\alpha}}  \simeq 98 \Lambda^{-1}$ given by Eq.(\ref{eq:k_c_1d}), see the vertical dashed red lines.
Accordingly, the IS evolves in the strong nonlinear regime since the largest frequency verifies $k_c \sim \xi^{-1}$, see Eq.(\ref{eq:corr_length_IS}).
}
\label{fig:NLSvsVlasov} 
\end{figure}
\end{center}

In this section we study the properties of the IS spontaneously generated in SPE simulations through the analysis of stationary solutions to the 1d WT-VPE (\ref{eq:vlasov_N0}-\ref{eq:V_0_inc}) -- note that the three-dimensional case $d=3$ is considered in section~\ref{sec:discussion} and Appendix~\ref{app:F}.
For this purpose we follow the general procedure reported in Ref.\cite{binney}. 
We will see that this type of stationary solutions needs to be adapted to our problem in order to compare it to the numerical simulations of the original SPE with a random wave $\psi(x,t)$.
We anticipate that a reasonable agreement is obtained between the stationary solution and the simulations (see Fig.~\ref{fig:NLSvsVlasov}).
Also note that the structure and the formation of incoherent halos have been widely studied through 
Lynden-Bell statistical equilibrium distributions \cite{lynden_bell67,chavanis-sommeria-98,chavanis98,RuffoPR,ruffo_book}, or through alternative correlation function approaches \cite{niemeyer18}.

The starting point is the observation that any stationary solution of the long-range WT-VPE (\ref{eq:vlasov_N0}-\ref{eq:V_0_inc}) can be expressed as an arbitrary function of the reduced Hamiltonian $h=\frac{\alpha}{2}k^2+V_{st}(x)$.
In the literature, there exists a large variety of equilibria solutions for collisionless systems, as broadly documented in \cite{binney}.
Our purpose here is to follow a natural procedure aimed at explaining the nature of the compact support of the spectrum observed in the SPE simulations, see Fig.~\ref{fig:inc_sol_regim_1d}(e).
The idea of the method is to argue that `particles' that constitute the stationary IS, with average density $n_{st}(h)$, are trapped by the self-consistent potential $V_{st}(x)$ 
(supposed to reach its minimum at $x=0$) provided that their energy is negative $h \le 0$.
This determines a specific interval of momenta for the self-trapped particles $-k_c \le k \le k_c$, where $k_c=\sqrt{-2V_{st}(0)/\alpha}$ will be shown to correspond to the frequency cut-off of the compactly supported spectral shape observed in the SPE simulations.

According to Eq.(\ref{eq:N_0}), we have $N_{st}(x)=(2\pi)^{-1}\int_{-k_c}^{k_c} n_{st}(k,x) dk$, and using a change of variable ($k \to h$) we get
\begin{eqnarray}
N_{st}(x)=\frac{1}{\pi \sqrt{2 \alpha}} \int_{V_{st}(x)}^0 \frac{n_{st}(h)}{\sqrt{h-V_{st}(x)}} dh,
\label{eq:N_st_cons}
\end{eqnarray}
Following a polytrope model \cite{binney}, we look for a solution in the form 
\begin{eqnarray}
n_{st}(h)=d_0 |h|^{p-1/2},
\label{eq:stat_sol_vlasov}
\end{eqnarray}
where $d_0$ is a constant and $p > 1$.
By integration of (\ref{eq:N_st_cons}) we get
\begin{eqnarray}
N_{st}(x)=r_p \big(-V_{st}(x)\big)^{p},
\label{eq:NV}
\end{eqnarray}
where $r_p=\frac{d_0}{\sqrt{2\pi \alpha}} \frac{ \Gamma(p+1/2)}{\Gamma(p+1)}$, $\Gamma(x)$ being the Gamma function.
Recalling the Poisson equation verified by the incoherent potential $\partial^2_x V_{st}(x)=2\gamma N_{st}(x)$, taking the second-order derivative of (\ref{eq:NV}) leads to the following form of the Emden equation for the stationary IS $N_{st}(x)$:
\begin{eqnarray}
\partial^2_x  N_{st}^{\frac{1}{p}}(x) =-2 \gamma r_p^{\frac{1}{p}} N_{st}(x).
\label{eq:emden_N_0_1d}
\end{eqnarray}
It proves convenient to recast Eq.(\ref{eq:emden_N_0_1d}) in the generic form of the Emden equation.
This can be done by introducing the spatial variable ${\bar x}=\sqrt{\eta_p}x$ and amplitude ${u}({\bar x})=(N_{st}({\bar x}/\sqrt{\eta_p})/N_{st}(0))^{\frac{1}{p}}$, with $\eta_p=2\gamma r_p^{\frac{1}{p}} N_{st}^{1-\frac{1}{p}}(0)$ and $u({\bar x}) \ge 0$.
Eq.(\ref{eq:emden_N_0_1d}) then recovers the standard form of the Emden equation
\begin{eqnarray}
\partial^2_{{\bar x}} u({\bar x})=-u^p({\bar x}),
\label{eq:emden}
\end{eqnarray}
with $u(0)=1$ and $\partial_{\bar x}u(0)=0$.
The Emden Eq.(\ref{eq:emden}) is known to exhibit a family of solutions that depend on the free parameter $p$.
These solutions are compactly supported in $[-{\bar x}_p,{\bar x}_p]$ for some $\bar{x}_p>0$ for any $p >1$.
In the following we denote the integral $F_{1,p}=\int_0^{{\bar x}_p} u^p({\bar x}) d{\bar x}$, where 
$u(\bar{x})$ is the solution to the one-dimensional Emden Eq.(\ref{eq:emden}).

The stationary IS is characterized by the mass 
\begin{eqnarray}
{\cal M} = g_{\cal M}(p)  F_{1,p}  N_{st}^{ \frac{p+1}{2p} } (0) ,
\label{eq:calN_1d} 
\end{eqnarray}
where 
{$g_{\cal M}(p)=(2 / \gamma )^{\frac{1}{2}}\big(\sqrt{2 \pi \alpha}/d_0\big)^{\frac{1}{2p}}\big( \frac{\Gamma(p+1)}{\Gamma(p+\frac{1}{2})}\big)^{\frac{1}{2p}}$}, 
by the linear contribution to the Hamiltonian ${\cal H}_l = \frac{\alpha}{4\pi} \iint k^2 n_{st}(k,x) dk dx$,
\begin{eqnarray}
{
{\cal H}_{l} = 
g_{{\cal H}}(p) \frac{F_{1,p+1}}{p+1}  N_{st}^{ \frac{p+3}{2p} } (0)  ,
}
\label{eq:calH_lin_1d} 
\end{eqnarray}
where $g_{{\cal H}}(p)=(2 \gamma )^{-\frac{1}{2}}\big(\sqrt{2 \pi \alpha}/d_0\big)^{\frac{3}{2p}}\big( \frac{\Gamma(p+1)}{\Gamma(p+\frac{1}{2})}\big)^{\frac{3}{2p}}$, 
{and by the nonlinear contribution ${\cal H}_{nl}  = \frac{\gamma}{2} \iint |x-y| N_{st}(x) N_{st}(y) dx dy$:
\begin{equation}
{\cal H}_{nl} = g_{{\cal H}}(p) \frac{G_{1,p}}{4} N_{st}^{ \frac{p+3}{2p} } (0),
\label{eq:calH_nl_1d}
\end{equation}
where
$G_{1,p}=\iint_{[-{\bar x}_p,{\bar x}_p]^2}|\overline{x}-\overline{y} |u(\overline{x})^p u(\overline{y})^p d\overline{x}d\overline {y}$.
We have 
(see Appendix \ref{app:D}):
\begin{equation}
\label{eq:relieFpGp}
\frac{F_{1,p+1}}{p+1} =\frac{G_{1,p}}{8}
\end{equation}
for all $p >1 $,
so that we find ${\cal H}_{nl}=2{\cal H}_l$.
This is in agreement with the virial theorem, which provides a general relation between the moment of inertia $I(t)=\int x^2 N(x,t) dx$ and the linear and nonlinear contributions to the Hamiltonian of the Vlasov equation
(\ref{eq:vlasov_N0}-\ref{eq:V_0_inc}).
Considering the WT-VPE for $d=1$, the virial theorem states that in a stationary state, $\partial_{t}^2 I=0$ and the energies should verify ${\cal H}_{nl}=2 {\cal H}_{l}$, see Appendix~\ref{app:E}.
Since ${\cal H}={\cal H}_l+{\cal H}_{nl}$ is preserved, we have
${\cal H}_{l}={\cal H}/3$ and ${\cal H}_{nl}=2{\cal H}/3$ with
\begin{equation}
{\cal H} = g_{{\cal H}}(p) \frac{3 F_{1,p+1}}{p+1}  N_{st}^{ \frac{p+3}{2p} } (0) .
\label{eq:calH_tot_1d}
\end{equation}
Note finally that  $V_{st}$ is not equal to the potential $V = - \gamma U_1*N_{st} $, they are only equal up to an additive constant
$V(x)-V_{st}(x) = V(0)-V_{st}(0)$, with
\begin{align}
\label{eq:V0_1d} 
V(0) &=g_{V}(p) H_{1,p} N_{st}(0)^{\frac{1}{p}}  , \\
V_{st}(0)  &=- g_{V}(p) N_{st}(0)^{\frac{1}{p}} , 
\end{align}
where $g_V(p) =\big(\sqrt{2 \pi \alpha}/d_0\big)^{\frac{1}{p}}\big( \frac{\Gamma(p+1)}{\Gamma(p+\frac{1}{2})}\big)^{\frac{1}{p}}$ 
and $H_{1,p} = \int_0^{{\bar x}_p} \bar{x} u^p(\bar{x})d\bar{x} $.
}

\subsection{The incoherent structure in the numerical simulations}
\label{subsec:ISnum}
In the previous subsection we have exhibited a family of theoretical stationary solutions of the  WT-VPE that can be parameterized by the values of $N_{st}(0)$, $d_0/N_{st}(0)$, and $p$.
It turns out that the numerical simulations of the SPE that are carried out in this paper produce solutions whose ISs tend to converge to these stationary solutions. 
The parameters of these stationary solutions 
can be determined from the initial conditions of the numerical simulations.
Indeed, the initial condition of the simulations is uniform in the box $[-L/2,L/2]$ with density $\bar{\rho}$, so that the numerical mass and Hamiltonian are ${\cal M}=\bar{\rho} L$ and ${\cal H}= \gamma {\cal M}^2 L / 8$ (because ${\cal H}_{nl,L} (t=0)= {\cal H}_l(t=0)=0$).
Once a quasi-equilibrium stationary state has been reached we have ${\cal H}_{l} = {\cal H}/3$ and ${\cal H}_{nl} = 2 {\cal H}/3$, so that 
 \begin{equation}
 \label{eq:Hlnltheo}
 {\cal M}=\bar{\rho}L,\quad \quad 
 {\cal H}_{l} =  \frac{\gamma \bar{\rho}^2 L^3}{24},
 \quad\quad
 {\cal H}_{nl,L} = - \frac{\gamma \bar{\rho}^2 L^3}{24}.
 \end{equation}
 Eqs.~(\ref{eq:calN_1d}) and (\ref{eq:calH_lin_1d}) then give two independent
 equations to determine the three parameters $N_{st}(0)$, $d_0/N_{st}(0)$, and $p$.
Note that Eqs.~(\ref{eq:calH_nl_1d}) and (\ref{eq:calH_tot_1d}) are equivalent to  (\ref{eq:calH_lin_1d}) and cannot be used as a third independent equation.
It is, however, possible to get a third independent equation  by using an additional hypothesis, that is verified numerically, which is that the support of the stationary incoherent structure essentially spans the box $[-L/2,L/2]$.
By using this hypothesis, the analysis in the Appendix~\ref{app:D} shows that the parameter $p$ is equal to 
\begin{equation}
\label{eq:valuep}
 p \simeq 4.76,
\end{equation}
which is solution of the algebraic equation (\ref{eq:alge}),
and the parameter $N_{st}(0)$ is given by
\begin{eqnarray}
 N_{st}(0) &=& \frac{6 F_{1,p+1}}{(p+1)F_{1,p}^3} \bar{\rho} \simeq  2.72  \bar{\rho}  .
\label{eq:valueNst0}
\end{eqnarray}
In addition, there is an important quantity that characterizes the stationary IS, namely the frequency cut-off $k_c$ of the spectrum that is compactly supported on $[-k_c,k_c]$ with $k_c=\sqrt{2/\alpha}(-V_{st}(0))^{1/2}$:
\begin{equation}
k_c =
\frac{\sqrt{\gamma \bar{\rho}} L}{\sqrt{\alpha}} \Big( \frac{F_{1,p} (p+1)}{6 F_{1,p+1}}\Big)^{\frac{1}{2}} \simeq
1.03 \, \frac{\sqrt{\gamma \bar{\rho}} L}{\sqrt{\alpha}}
.
\label{eq:k_c_1d}
\end{equation}
The frequency cut-off $k_c$ has a simple physical meaning, namely that `particles' with a momentum $|k| \le k_c$ are characterized by a negative energy $h=(\alpha/2)k^2+V_{st}(x) \le 0$, so that they get trapped by the self-induced potential $V_{st}(x)$ of the IS.
This means that linear and nonlinear effects balance each other and are thus of the same order of magnitude: using the definition of the healing length $\xi$ in (\ref{eq:healing_length}), the (smallest) correlation length of the IS reads 
\begin{equation}
\lambda_c \sim 1/k_c \sim \xi,
\label{eq:corr_length_IS}
\end{equation}
as anticipated above through the separation of spatial scales (\ref{eq:sep_scales}).
This confirms that the IS does not evolve in the weakly nonlinear (i.e. kinetic) regime described by the weak turbulence kinetic equation \cite{nazarenko_nse}.

We have compared the results of the numerical simulations of the SPE with the stationary solution of the long-range WT-VPE discussed in this section.
We recall that the simulations start from a homogeneous initial condition, which exhibits the gravitational instability and eventually the formation of the localized IS, as discussed above through Fig.~\ref{fig:inc_sol_regim_1d}.
The convergence of the numerical values of  ${\cal H}_{l} $ and ${\cal H}_{nl,L}$ to the theoretical values (\ref{eq:Hlnltheo}) can be observed in Fig.~\ref{fig:inc_sol_regim_1d}(d) (see the dashed horizontal lines). 
At this stage we can compare the IS obtained in the SPE simulations with the stationary solution of the WT-VPE in which the parameters $p$ and $N_{st}(0)$ are given by (\ref{eq:valuep}-\ref{eq:valueNst0}).
The comparison is reported in Fig.~\ref{fig:NLSvsVlasov} for a small value of the parameter ${\tilde \xi} \simeq  1.5 \times 10^{-2}$ (corresponding to Fig.~\ref{fig:inc_sol_regim_1d}).
We can note that a good agreement has been obtained between the SPE simulations and the stationary solution of the WT-VPE -- note however that such a good agreement is not obtained in the tail of the localized IS due to the periodic boundary conditions.
On the other hand, we observe a good agreement for the spectrum in logarithmic scale and the corresponding frequency cut-off $k_c$ given by Eq.(\ref{eq:k_c_1d}). 
We note that the good agreement shown in Fig.~\ref{fig:NLSvsVlasov} has been also obtained with a variety of different numerical parameters (in the regime ${\tilde \xi} \ll 1$).

\section{Stabilization of the coherent soliton by the incoherent structure}
\label{sec:stabilization}

The effective SPE (\ref{eq:nls_A0}) has revealed that the IS introduces an effective parabolic potential that confines the underlying coherent soliton.
In this section we study the stabilization of such a hidden soliton by the surrounding IS.
For this purpose, we have performed numerical simulations of the system of coupled SPE and WT-VPE given in (\ref{eq:nls_0}-\ref{eq:N_0}).
It is important to note that these coupled equations for the coherent and incoherent components have been derived directly from the original SPE with rather weak assumptions.
In the following we show that a stationary solution for the soliton 
is robust when substituted into the original system of coupled SPE and long-range WT-Vlasov Eqs.(\ref{eq:nls_0}-\ref{eq:N_0}).

\begin{center}
\begin{figure}
\includegraphics[width=1\columnwidth]{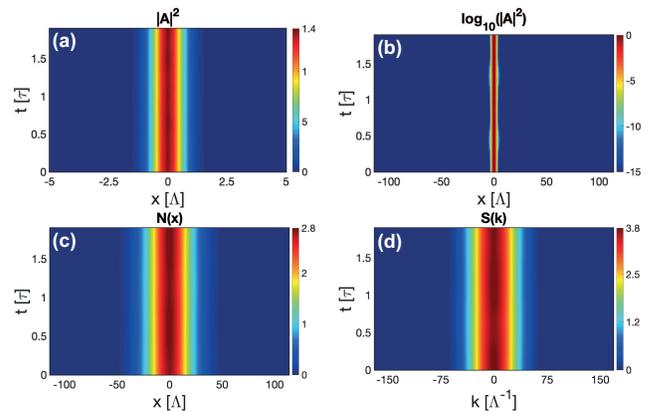}
\caption{
\baselineskip 10pt
{\bf Dynamics of the soliton in the presence of the IS: The soliton is stabilized by the IS.}
Numerical simulation of the coupled system SPE and WT-VPE (\ref{eq:nls_0}-\ref{eq:N_0}) for $L=114 \Lambda$ and ${\tilde \xi}\simeq 1.75 \times 10^{-2}$. 
Evolution of the soliton density $|A|^2(x,t)$ in normal scale (a), logarithmic scale (b).
Evolution of the density of the IS $N(x,t)=\frac{1}{2\pi}\int n(k,x,t) dk$ (c) and corresponding spectrum $S(k,t)=
\int n(k,x,t) dx$ (d), starting from the stationary solution of the WT-VPE (\ref{eq:stat_sol_vlasov}), see section~\ref{sec:stat_sol_vlasov_1d}. 
The initial condition for the soliton corresponds to the stationary Gaussian-shaped solution Eq.(\ref{eq:stat_sol_1d}).
As predicted by the theory, the IS provides a spatial confinement that stabilizes the Gaussian-shaped soliton, as evidenced by the corresponding evolution of the soliton {\it in the absence of the IS}, see Fig.~\ref{fig:sim_nls_no_vlasov}.
The densities $|A(x)|^2$ and $N(x)$ are in units of ${\bar \rho}$.
}
\label{fig:sim_nls_vlasov} 
\end{figure}
\end{center}

We have performed numerical simulations of the coupled SPE and long-range WT-VPE (\ref{eq:nls_0}-\ref{eq:N_0}) in one spatial dimension, $d=1$, see Fig.~\ref{fig:sim_nls_vlasov}.
The initial conditions for the coherent soliton and the IS have been chosen as follows.
For the coherent soliton component, we consider as initial condition $A(x,t=0)$ the stationary soliton solution that accounts for the presence of the surrounding IS, see Ref.\cite{PRL21}: 
\begin{eqnarray}
|A|^2(x,t=0)=a^2 \exp\Big(-\frac{x^2}{2 R_S^2}\Big),
\label{eq:stat_sol_1d}
\end{eqnarray}
where the mass $M_S=\int |A|^2 dx=a^2\sqrt{\pi}R_S$ is related to the soliton radius by the relation
\begin{eqnarray}
M_{S} = \frac{\sqrt{\pi}}{\sqrt{2}\gamma} \Big( \frac{\alpha}{R_S^3}-2\gamma N_{st} (0) R_S \Big).
\end{eqnarray}
For the IS governed by the WT-VPE (\ref{eq:vlasov_0}), the initial condition $n(k,x,t=0)$ corresponds to the stationary solution obtained in section~\ref{sec:propertiesIS}.
In this example the peak density of the soliton has been chosen to be equal to $|A|^2(x=0,t=0)=N_{st}(0)/2$.
The numerical simulation of the  coupled SPE and WT-VPE (\ref{eq:nls_0}-\ref{eq:N_0}) shows that the IS remains almost stationary during the evolution (except for a small readjustment to the constant homogeneous background added to the stationary solution in the phase-space region $h=(\alpha/2)k^2+V_{st}(x) > 0$ where the solution is not defined).
This means that the presence of the soliton does not affect the evolution of the stationary IS, as predicted by the scaling developed in section~\ref{sec:scaling}.
Along this line, the coherent soliton $A(x,t)$ governed by Eq.(\ref{eq:nls_0}) also remains almost stationary during the evolution, except for a small oscillation in the far tail ($\sim 10^{-10}$) of the Gaussian density profile.

\begin{center}
\begin{figure}
\includegraphics[width=1\columnwidth]{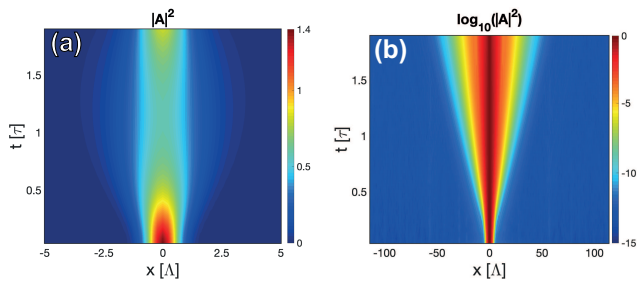}
\caption{
\baselineskip 10pt
{\bf Dynamics of the soliton in the absence of the IS.}
Evolution of the soliton density $|A|^2(x,t)$ in normal (a), and logarithmic scale (b), obtained by simulation of the SPE for the same parameters and same initial condition as in Fig.~\ref{fig:sim_nls_vlasov}, but without any coupling to the IS.
The soliton exhibits a rapid and significant broadening of its spatial profile, which is in contrast with the stationary dynamics reported in Fig.~\ref{fig:sim_nls_vlasov} {\it in the presence of the IS}.
The density $|A(x)|^2$ is in units of ${\bar \rho}$.
}
\label{fig:sim_nls_no_vlasov} 
\end{figure}
\end{center}

This evolution of the soliton is in marked contrast with the corresponding dynamics obtained {\it in the absence of the IS}.
This is illustrated in Fig.~\ref{fig:sim_nls_no_vlasov}, that shows the evolution of the soliton density $|A|^2(x,t)$ starting from the same initial condition as in Fig.~\ref{fig:sim_nls_vlasov}, but {\it without the IS}, i.e. by setting $N(x,t)=0$ in the potential $V(x,t)$ in Eq.(\ref{eq:V_0}).
In this case the soliton exhibits its self-gravitating potential and its evolution is featured by a rapid and significant broadening of the localized structure, as evidenced by the logarithmic plot of the density $|A|^2(x,t)$ reported in Fig.~\ref{fig:sim_nls_no_vlasov}.
The comparison of this evolution with that obtained in Fig.\ref{fig:sim_nls_vlasov} clearly shows that the IS provides a spatial confinement that stabilizes the Gaussian-shaped soliton.
We note that the same evolution of the coherent soliton reported in Fig.~\ref{fig:sim_nls_vlasov} in the presence of the IS is also obtained by numerical simulation of the effective SPE (\ref{eq:nls_A0}) alone, in which the IS introduces a parabolic confinement for the soliton.
These numerical simulations then provide a confirmation of the theoretical developments.


\section{Discussion and extensions}
\label{sec:discussion}

In summary, by decreasing the normalized healing length ${\tilde \xi}$ (or the parameter $\Xi$ in \cite{mocz17}), the long-range wave system described by the SPE enters an incoherent collective regime characterized by the formation of a large scale IS that contains hidden solitons.
Although the soliton can hardly be identified  in the usual spatial or spectral domains, its presence is revealed by a phase-space analysis of the wave amplitude.
We have derived a set of coupled equations describing the evolutions of the soliton component governed by an effective SPE equation and the IS component governed by the WT-VPE.
In this way, we have obtained a stationary solution of the WT-VPE that has been found in  agreement with the numerical simulations of the SPE, in particular by describing the {\it origin of the compact support of the wave spectrum observed in the simulations}.
The theory reveals that the IS component introduces an effective parabolic potential that confines and helps stabilizing the coherent soliton.
As a consequence, the soliton exhibits a Gaussian density profile and the corresponding stationary solution has been verified by numerical simulations of the coupled SPE and long-range WT-VPE system of equations.
The typical width of the soliton is much larger than the correlation length of the IS and much smaller than the typical width of the IS, see Eq.(\ref{eq:sep_scales}).

In the following we discuss the natural extension of our theory to three spatial dimensions $d=3$.
This clarifies the quantum-to-classical (SPE to VPE) correspondence $\hbar/m \to 0$: By combining the WT-VPE stationary solution with the mass-radius relation for the coherent soliton, we obtain the following scaling for the soliton radius $R_S/L \propto {\tilde \xi}$ and for the soliton mass $M_S/M_{IS} \propto {\tilde \xi}^3$. 
We also report in the Appendix~\ref{app:G} numerical simulations of the SPE in two spatial dimensions ($d=2$) for moderate values of the parameter ${\tilde \xi}$ accessible in our simulations, which evidence the formation of a partially hidden soliton.
In addition, we show below that the regime of `hidden' solitons is not restricted to the gravitational interaction, and can be extended to long-range interactions characterized by an algebraic decay of the nonlocal nonlinear potential.
Finally, we discuss possible experimental implementations of the predicted regime of hidden coherent soliton states.

\subsection{Extension to three spatial dimensions: The quantum-to-classical correspondence}


The regime of hidden solitons can be extended to three spatial dimensions \cite{PRL21}.
By decreasing the parameter ${\tilde \xi}$, the system should enter the collective incoherent regime characterized by the formation of a large scale IS (described by the WT-VPE (\ref{eq:vlasov_N0}-\ref{eq:V_0_inc})), which should hide and sustain a coherent soliton state (described by the effective SPE (\ref{eq:nls_A0}-\ref{eq:V_0_coh})).
In this section, we combine the stationary solutions for the IS with the mass-radius relation for the coherent soliton to obtain a more precise relation for the soliton radius and soliton mass vs the normalized healing length, which clarifies the quantum-to-classical correspondence in 3d.

First of all, we make use of the variational approach applied to the effective SPE (\ref{eq:nls_A0}-\ref{eq:V_0_coh}) to describe the generic oscillatory dynamics of a Gaussian-shaped soliton, see the Appendix~\ref{sec:stat_sol_CS_3d} or Ref.\cite{PRL21}.
The analysis reveals in particular the existence of a stable stationary 3d soliton featured by a Gaussian profile 
\begin{equation}
|A|^2(\bx)=a^2 \exp\Big(- \frac{|\bx|^2}{2 R_S^2}\Big),
\end{equation} 
where the mass $M_{S}=\pi^{3/2} a^2 R_S^3$ and the soliton radius are related by
\begin{equation}
M_{S} = \frac{\sqrt{2\pi}}{\gamma}\Big( \frac{3\alpha}{2R_S}- 2 \pi \gamma N_{st} ({\bf 0}) R_S^3\Big)  .
\label{eq:masse_3d}
\end{equation}
The solution $R_S(M_{S})$ is a decaying function of $M_{S}$, that goes from $R_{\rm max} = \big(3\alpha/(4\pi \gamma N_{st} ({\bf 0})) \big)^{1/4}$ for vanishing $M_{S}$ to $0$ when $M_{S}\to +\infty$.
The amplitude $N_{st} ({\bf 0})$ of the IS in (\ref{eq:masse_3d}) is obtained form the stationary solution of the WT-VPE (\ref{eq:vlasov_N0}-\ref{eq:V_0_inc}).
The reader can find the details of the derivation of 3d stationary solution in the Appendix~\ref{sec:stat_sol_vlasov_3d}.
In particular, the 3d IS is characterized by a spectrum with a compact support $|\bk| \in (0, k_c)$ with the frequency cut-off  $k_c$ given by Eq.(\ref{eq:kc_3d}).

According to the correspondence between the SPE and the kinetic Vlasov-Poisson (collisionless Boltzmann) equation, one should expect that soliton states should disappear in the classical limit $\hbar/m \to 0$, since they are genuine coherent objects that cannot be described by the classical limit \cite{mocz18}.
We discuss this aspect through the regime of hidden coherent solitons featured by a Gaussian-shaped density profile (instead of the usual algebraic soliton profile without the trapping potential provided by the IS).
For this purpose, let us consider Eq.(\ref{eq:masse_3d}) expressed in term of the peak density $a^2=M_{S}/(\pi^{3/2} R_S^3)$ of the soliton instead of the mass.
Following the multiple scale analysis for ${\tilde \xi}\ll 1$ [see Eqs.(\ref{eq:A_0}-\ref{eq:n_0})], the soliton density is of the same order as the density of the IS, then we pose $a^2=\eta N_{st}({\bf 0})$ with $\eta \sim 1$ to get
\begin{equation}
R_S^4 = \frac{3\alpha}{\sqrt{2}\pi (\eta+2^{3/2}) \gamma N_{st}({\bf 0})  }.
\label{eq:corresp_3d_0}
\end{equation} 
Next we make use of the assumption that the initial condition of the field is homogeneous, so that 
we get an explicit expression for $N_{st}({\bf 0})$, see Eq.(\ref{eq:N_0vsp_3d}) in Appendix~\ref{app:F}.
This gives the following closed relation for the three-dimensional radius of the soliton
\begin{equation}
\frac{R_S}{L} = \varrho^{\frac{1}{4}} \Big(  \frac{12\pi  F_{3,p+1} }{ (p+1) c_3 }   \Big)^{\frac{3}{4}} F_{3,p}^{-\frac{5}{4}} \ {\tilde \xi},
\label{eq:corresp_3d}
\end{equation}
where the prefactor $\varrho=3 \sqrt{2}/(\pi(\eta+2^{3/2})) \sim 1$.
We can therefore expect $R_{s}/L \to 0$ as ${\tilde \xi} \to 0$, i.e., the soliton should vanish consistently with the kinetic Vlasov-Poisson correspondence.
The same conclusion is obtained from the analysis of the mass of the Gaussian-shaped soliton:
\begin{equation}
\frac{M_{S}}{M_{IS}} = \eta \varrho^{\frac{3}{4}} \Big(  \frac{\pi (p+1) c_3}{12\pi F_{3,p+1}} \Big)^{\frac{3}{4}} F_{3,p}^{\frac{5}{4}}  {\tilde \xi}^3,
\label{eq:corresp_3d_scale}
\end{equation}
where $M_{IS}$ is the mass of the IS, see Eq.(\ref{eq:rapp_masses}).
Note that the scaling ${M_{S}}/{M_{IS}} \sim
{\tilde \xi}^3 \sim \Xi^{3/4}$ is close but not equal to the scaling $\Xi^{2/3}$ predicted in \cite{mocz18} for a given halo cuspy profile.
We finally remark that, although the soliton disappears in the purely classical limit ${\tilde \xi} \to 0$, for any small, yet positive, value of ${\tilde \xi}$, the soliton still survives in the form of a Gaussian-shaped coherent structure with peak density of the same order as the average density fluctuations of the IS, $|A|^2 \sim N_{st} ({\bf 0})$ and a typical radius $R_S$ given by the harmonic average of $\xi$ and $L$.

\subsection{Extension to non-gravitational long-range potentials}

We have considered in this paper the concrete examples of gravitational potentials in different spatial dimensions.
However it is important to note that the regime of hidden coherent solitons described here is general and can be extended to generic long-range potentials.
We illustrate this in one spatial dimension by considering a general form of the nonlinear Schr\"odinger Eq.(\ref{eq:nse_0}) and (\ref{eq:nse_3}) with the following  potential that decays algebraically in space
\begin{eqnarray}
U_\nu(x)=  -|x|^\nu, \quad \nu \in (0,1). 
\label{eq:U_0}
\end{eqnarray}

Following the same procedure reported in Ref.\cite{PRL21},  the multiple scale series expansion provides the two rescaled functions
\begin{eqnarray}
A(x,t) = A^{(0)}(x,t),
\label{eq:scaling_A0_1d}\\
n(k,x,t) = \eps^\nu n^{(0)}(\eps k, \eps x, t).
\label{eq:scaling_n0_1d}
\end{eqnarray}
The coherent soliton component satisfies an effective SPE
\begin{eqnarray}
&& i\partial_t A  = - \frac{\alpha}{2} \partial_{xx}  A + V_{S} A + \gamma  Q(t) |x|^2 A \quad
\label{eq:nls_A0_r} \\
&& Q(t) =- \nu \int_0^\infty y^{\nu-1} \partial_y N(y) dy , \\
&& V_{S}(x,t)= - \gamma \int U_\nu(x-y) |A|^2(y,t) dy,
\label{eq:V_0_coh_r} 
\end{eqnarray}
while the incoherent component verifies a rescaled WT-VPE
\begin{eqnarray}
&& \partial_t n(k,x) +\alpha k  \partial_{x} n(k,x) 
- \partial_{x} V_{IS}  \partial_{k} n(k,x) = 0 
\quad \quad
\label{eq:vlasov_N0_r} \\
&& V_{IS}(x,t) = - \gamma \int U_\nu(x-y)N(y,t) dy.
\label{eq:V_0_inc_r}
\end{eqnarray}
These rescaled SPE and WT-VPE generalize those derived above for the particular gravitation potential $\nu=1$ (note in particular that $Q(t) =N(0,t)$ for $\nu=1$).
In particular, by following the same procedure, a stable soliton solution of the effective SPE (\ref{eq:nls_A0_r}) can be found by the Lagrangian approach, which gives a relation between the mass of the coherent structure $M_{S}=a^2 \sqrt{\pi} R_S$ and the width of the Gaussian-shaped profile $|A|^2(x)=a^2 \exp(-x^2/(2 R_S))$:
\begin{equation}
M_{S}=\frac{1}{2\nu\gamma c_\nu} \Big( \frac{\alpha}{R_S^{2+\nu}} -2 \gamma Q R_S^{2-\nu}  \Big),
\end{equation}
where the amplitude $Q$ can be obtained from the stationary solution of the WT-VPE (\ref{eq:vlasov_N0_r}-\ref{eq:V_0_inc_r}), as discussed above for $\nu=1$.

This multi-scale analysis reveals a major difference with respect to the gravitational potential $\nu=1$.
Since $N(x,t) = \eps^{\nu-1} N^{(0)}(\eps x,t)$, with $N^{(0)}(\eps x,t)=\frac{1}{2\pi}\int n^{(0)}(\eps x,\eps k,t)d(\eps k)$, then the soliton density $|A|^2$ is smaller than the average density $N(x)$ of the fluctuations of the IS, i.e., the soliton becomes `more hidden' than in the gravitational case $\nu=1$.
As discussed above the presence of the soliton can be identified in the phase-space representation.
Indeed, the soliton appears as  a peak in phase-space of amplitude of order one compared to the background $n(k,x)$ whose amplitude is of order $\eps^\nu$.
Accordingly, for a moderate value of $\nu \sim 0.5$, the identification of the soliton can become difficult  even through the phase-space analysis. 

\subsection{Possible experimental implementations}

This work should stimulate experiments in different fields.
We mention the example of dipolar Bose-Einstein condensates, which are known to exhibit a genuine long range nonlocal potential \cite{dipolarBEC}.
In the context of nonlinear optics, a nonlocal nonlinearity \cite{krolikowski04} can be found in atomic vapours \cite{vapor}, nematic liquids crystals \cite{peccianti04,NLCliqcryst,conti04,Residori12}, or thermal nonlinear media \cite{Segev_rev,kivshar_agrawal03,PR14,krolikowski04,rotschild06,
ghofraniha07,cohen06,rotschild08,rotschild05,vocke16,marcucci19}.
Optical beam propagation in the nonlinear medium can be described in the paraxial approximation by a generic form of the nonlocal nonlinear Schr\"odinger Eq.(\ref{eq:nse_0})-(\ref{eq:nse_3}), where the variable $t$ plays the role of the propagation distance in the medium and the remaining spatial directions are transverse to the beam. 
Along this line, a formal analogy has been pointed out between the SPE Eq.(\ref{eq:nse}-\ref{eq:poiss}) for $d=2$ and the nonlocal nonlinear Schr\"odinger equation governing light propagation in highly nonlocal nonlinear media, which allowed to emulate a 2d slice of the full 3d gravitational system \cite{segev15,faccio_bose_star}.
More precisely, choosing appropriate operating conditions, it has been shown that the medium response in the transverse plane of the optical beam can be made to mimic the long-range nature of the gravitational potential.
This procedure allowed for the experimental study of the evolution of a rotating Bose star with quantized angular momentum using intense light beam propagating in a lead-glass slab \cite{faccio_bose_star}.
Following this experimental procedure, we can envisage the experimental observation of the regime of hidden coherent soliton states reported in this work.
It is important to note in this respect that the hidden solitons can be unveiled experimentally
through the measurement of the optical spectrogram, which provides the phase-space representation (Husimi transform) of the coherence properties of an optical beam \cite{waller12}, a feature already implemented experimentally in \cite{NC15}.
From a different perspective, it would  be interesting to study the long-term evolution of the system (i.e., much larger nonlinear interaction `times') in a confined wave-guided geometry by considering optical fibres filled with thermal nonlinear liquids \cite{bang17}, in relation with the recent experimental demonstrations of light thermalization and condensation in graded index multimode fibers \cite{wiseRJ,PRL19,PRL20}.

\section{Acknowledgements}

The authors are grateful to J. Niemeyer for drawing our attention to this problem and for the fruitful discussions and suggestions during the early stage of this work.
We acknowledge financial support from the French ANR under Grant No. ANR-19-CE46-0007 (project ICCI),
iXcore research foundation, EIPHI Graduate School (Contract No. ANR-17-EURE-0002), French program ``Investissement d'Avenir," Project No. ISITE-BFC-299 (ANR-15 IDEX-0003);  H2020 Marie Sklodowska-Curie Actions (MSCA-COFUND) (MULTIPLY Project No. 713694). Calculations were performed using HPC resources from DNUM CCUB (Centre de Calcul, Universit\'e de Bourgogne).


\bigskip

\appendix 

\begin{widetext}

\section{Numerical simulations of the SPE in a finite box}
\label{app:A}%
The numerical simulations are carried out on a system set in a bounded domain with prescribed boundary conditions.
In this paper we consider 
\begin{align}
\label{eq3}
i \partial_{{t}} {\psi} + \frac{\alpha}{2} \Delta_{{\bx}} {\psi} - {V}_L {\psi}  =0 , \\
\label{eq4}
 {V}_L = - \gamma U_{d,L} * |\psi|^2 ,
\end{align}
in $[-L/2,L/2]^d$ with periodic boundary conditions.
Here $U_{d,L}(\bx) $ is the periodic function equal to $U_d(\bx) -c_{d,L}$ in $[-L/2,L/2]^d$ and 
$c_{d,L} =  L^{-d} \int_{[-L/2,L/2]^d} U_d(\bx) d\bx$ so that $U_{d,L}$ has mean zero. 
Eq.~(\ref{eq4}) is equivalent to 
\begin{equation}
 {V}_L = - \gamma  U_{d} * ( |\psi|^2 - \bar{\rho}_L)  ,
 \end{equation}
 with $\bar{\rho}_L= L^{-d} \int_{[-L/2,L/2]^d} |\psi(\bx)|^2 d\bx$. 
Accordingly, the potential $V_L$ in the numerical simulations has zero mean 
and the corresponding nonlinear energy 
\begin{equation}
\label{def:HnlL}
{\cal H}_{nl,L}= \frac{1}{2}\int {V}_L(\bx,t) |\psi|^2(\bx,t) d\bx
\end{equation}
is reported in Figs.~\ref{fig:sol_regim_1d}, \ref{fig:inc_sol_regim_1d} with this potential. 
 This nonlinear energy is different from the nonlinear energy ${\cal H}_{nl}$ in Eq.(\ref{eq:H_nl}) by a time-independent constant
\begin{equation}
\label{def:HnlL_2}
{\cal H}_{nl,L}= {\cal H}_{nl} +\gamma {\cal M}^2 c_{d,L}/2,
\end{equation}
where ${\cal M}= \int_{[-L/2,L/2]^d} |\psi(\bx)|^2 d\bx$.
The potential $V_L$ is numerically computed by Fourier transform \cite{mocz17}.
Note that the formulations (\ref{eq:nse_0}-\ref{eq:nse_3}) in open medium (used for the theoretical analysis) and (\ref{eq3}-\ref{eq4}) in periodic medium (used for the numerical analysis) have an apparent departure in the definitions of the potential $U$ which differ by a constant. However, an additional constant in the potential can be removed from the SPE by means of a global phase change for $\psi$,
which does not affect the dynamics.


\section{SPE simulations for $d=2$}
\label{app:G}


The 1d simulations allowed us to numerically investigate a large separation of the three scales, $\lambda_c \sim \xi \ll R_S \ll \ell$ with small values of the normalized healing length ${\tilde \xi} \ll 1$, as discussed through (\ref{eq:sep_scales}).
Here we report simulations in $d=2$ for moderate small values of ${\tilde \xi}$, which qualitatively confirm the scenario discussed for $d=1$.

We have simulated the SPE with the two-dimensional gravitational potential $U_2(\bx)=-\log(|\bx|)$ by starting from a homogeneous initial condition with a super-imposed small noise, see Eq.(\ref{eq:U_d}) and Appendix~\ref{app:A}.
As for $d=1$, the system exhibits the gravitational (modulation) instability process, which is followed by a turbulence regime characterized by the interaction of solitons with the surrounding fluctuations.
For a moderate value of ${\tilde \xi}$ (typically ${\tilde \xi} \sim 0.3$), the quasi-stationary dynamics is characterized by the formation of a large amplitude soliton that dominates a surrounding small amplitude IS.
In addition, the typical radius of the soliton is of the order of the correlation length of the fluctuations of the incoherent halo.

By decreasing the parameter ${\tilde \xi}$, the system enters the collective incoherent regime characterized by the formation of a large scale IS. 
A typical example of this regime is reported in Fig.~\ref{fig:IS_CS_2d} for a moderate small value of ${\tilde \xi} \simeq 0.063$ that is accessible in our numerical simulations.
As for $d=1$, the evolutions of the energies ${\cal H}_l$ and ${\cal H}_{nl,L}$ reported in Fig.~\ref{fig:IS_CS_2d}(a) reflect a relaxation to a quasi-stationary state.
It is interesting to note that, at variance with $d=1$ or $d=3$, for $d=2$ the virial theorem fixes the stationary value of the linear energy exclusively from the system mass (independently of the particular initial condition), see Appendix~\ref{app:E}:
\begin{eqnarray}
{\cal H}_l=\gamma {\cal M}^2/4. 
\label{eq:H_l_viral2d}
\end{eqnarray}
In the simulation the initial condition is homogeneous, so that ${\cal H}_l(t=0)={\cal H}_{nl,L}(t=0)=0$, ${\cal M}={\bar \rho} L^2$ and by the conservation of the Hamiltonian the nonlinear energy of the IS in its stationary state is ${\cal H}_{nl,L}=-\gamma {\cal M}^2/4$.
These theoretical predictions of the virial theorem are found in agreement with the IS generated in 2d SPE simulations, as illustrated in Fig.~\ref{fig:IS_CS_2d}(a).

\begin{center}
\begin{figure}
\includegraphics[width=.5\columnwidth]{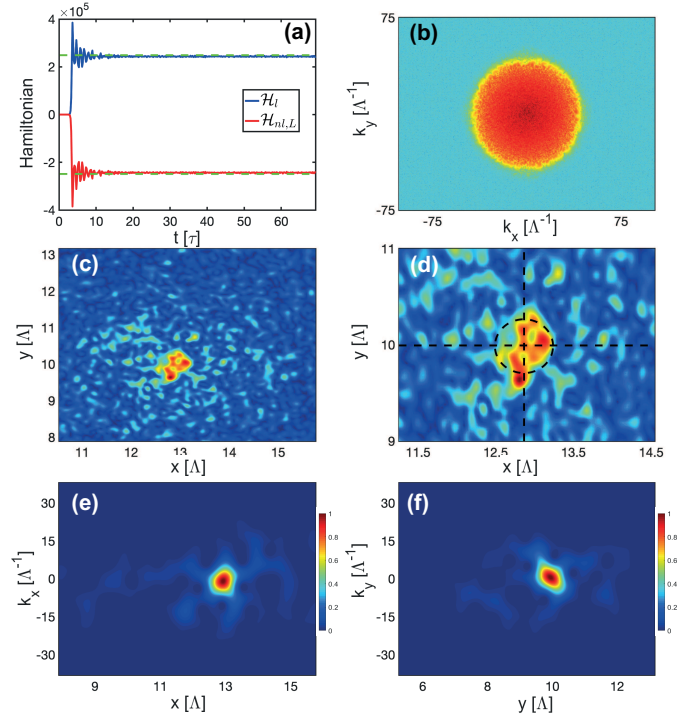}
\caption{
\baselineskip 10pt
{\bf Incoherent structure and partially hidden coherent soliton in two spatial dimensions ($d=2$).}
SPE simulation in a 2d box $[-L/2,L/2]^2$ of size $L=32 \Lambda$ with periodic boundary conditions for a moderate small value of ${\tilde \xi} \simeq 0.063$ 
starting from a homogeneous initial condition.
The initial wave develops the gravitational instability and after a transient the system self-organizes into a large scale incoherent localized structure that contains a partially hidden soliton.
(a) Evolutions of the linear ${\cal H}_l$ (blue) and nonlinear ${\cal H}_{nl,L}$ (red) energy contributions to the Hamiltonian.
The green dashed lines denote the theoretical values of ${\cal H}_l=\gamma {\cal M}^2/4$ and ${\cal H}_{nl,L}=-\gamma {\cal M}^2/4$, with ${\cal M}={\bar \rho} L^2$, predicted by the virial theorem once a quasi-stationary state has been reached, see Eq.(\ref{eq:H_l_viral2d}) and Appendix~\ref{app:E}.
(b) Two-dimensional spectrum $|{\tilde \psi}|^2(\bk)$ at $t=81 \tau$ in log$_{10}-$scale. Note that the spectrum exhibits a compact support.
(c) Two-dimensional density $|{\psi}|^2(\bx)$ at $t=81 \tau$.
(d) Zoom nearby the presence of the soliton.
Corresponding phase-space distributions $(x,k_x)$ (e), and $(y,k_y)$ (f), along the horizontal and vertical lines shown in (d): The spot in phase-space indicates the presence of a 2d soliton evidenced schematically in (d) by a circle.
Because of the moderate small value of ${\tilde \xi}$ considered in the simulation, we have $\lambda_c \lesssim R_S$: There is no clear separation of scales between the correlation length of the IS and the soliton radius (see Fig.~\ref{fig:inc_sol_regim_1d}(c) for comparison with $d=1$ where $\lambda_c \ll R_S$).
}
\label{fig:IS_CS_2d} 
\end{figure}
\end{center}

Furthermore, as for $d=1$, the IS is localized in space by its self-gravitating potential and it is characterized by a compactly supported spectral shape, see Fig.~\ref{fig:IS_CS_2d}(b).
Proceeding as for $d=1$, we look for a soliton hidden in the phase-space representation.
We report the Husimi transform along two representative $x$ and $y$ lines, which indicates the presence of a soliton whose width is larger than the correlation length of the fluctuations by a factor $\sim 3$.
This small factor is due to the moderate value of ${\tilde \xi} \simeq 0.063$ considered in the simulation (we recall that ${\tilde \xi}$ is computed with $\ell = L/2$ as for $d=1$).
This merely explains why the soliton is not fully `hidden' in the IS.
Unfortunately, because of the limited computational power, we are not able for $d=2$ to decrease the parameter ${\tilde \xi}$ to the small values considered for $d=1$ (i.e. ${\tilde \xi} \simeq 10^{-2}$).
For this reason, we do not reach for $d=2$ the required separation of scales $\lambda_c \ll R_S$ discussed in Eq.(\ref{eq:sep_scales}), as clearly revealed by the comparison of the phase-space portrait for $d=1$ (see Fig.~\ref{fig:inc_sol_regim_1d}(c)) and for $d=2$ (see Fig.~\ref{fig:IS_CS_2d}(f-g)).


\section{The incoherent structure for $d=1$}
\label{app:D}
\subsection{Mass  and energy of the incoherent structure}

Derivation of Eq.(\ref{eq:calN_1d}):
The mass ${\cal M}=\int N(x) dx=\frac{N_0}{\sqrt{\eta_p}} \int d{\bar x} u^p({\bar x})$. 
Using $\eta_p=2\gamma r_p^{1/p} N_{st}(0)^{1-1/p}$ and $r_p=\frac{d_0}{\sqrt{2\pi \alpha}} \frac{ \Gamma(p+\frac{1}{2})}{\Gamma(p+1)}$ gives after simplification the expression of the mass ${\cal M}$ in Eq.(\ref{eq:calN_1d}).

\bigskip

Derivation of Eq.(\ref{eq:calH_lin_1d}):
The linear Hamiltonian reads for $d=1$ \cite{PR14}
\begin{eqnarray}
{\cal H}_{l} = \frac{1}{2\pi} \iint dk dx  \frac{\alpha k^2}{2} n(k,x,t). 
\end{eqnarray}
Making the substitution $\frac{\alpha k^2}{2}=h-V_{st}$ and computing the integral over $k$ by proceeding as for the derivation of the stationary solution of the WT-VPE, we get
\begin{eqnarray}
{\cal H}_{l} 
=\frac{d_0}{2\sqrt{2\pi \alpha}} \frac{\Gamma(p+\frac{1}{2})}{\Gamma(p+2)} \frac{1}{r_p^{1+1/p}}\int dx N_{st}(x)^{1+1/p}.
\end{eqnarray}
where we made use of Eq.(\ref{eq:NV}).
Proceeding as here above for the mass ${\cal M}$, we obtain Eq.(\ref{eq:calH_lin_1d}). 

\bigskip

Derivation of Eq.(\ref{eq:calH_nl_1d}):
The nonlinear contribution of the Hamiltonian of the WT-VPE for $d=1$ is \cite{PR14}
\begin{eqnarray}
 {\cal H}_{nl} = -\frac{\gamma}{2} \iint dx dy U_1 (x-y) N_{st}(x)N_{st}(y)= \frac{N_{st}(0)^{\frac{p+3}{2p}}}{\sqrt{2\gamma} r_p^{\frac{3}{2p}}}
 \Big[ \frac{1}{4} \iint_{[-{\bar x}_p,{\bar x}_p]^2} |\bar{x}-\bar{y}| u(\bar{x})^p u(\bar{y})^p d\bar{x} d\bar{y}\Big],
\end{eqnarray}
which gives Eq.(\ref{eq:calH_nl_1d}).


\bigskip

Derivation of Eq.(\ref{eq:relieFpGp}):
On the one hand, by using the Emden equation (\ref{eq:emden}), by integrating by parts, and by using $\partial_{\bar{x}} u(0)=0$ and $u(\bar{x}_p)=0$, 
we have
\begin{equation}
\label{eq:appD4}
F_{1,p+1} = - \int_0^{\bar{x}_p} u \partial_{\bar{x}}^2 u d\bar{x} = - [u \partial_{\bar{x}} u]_0^{\bar{x}_p}  
+\int_0^{\bar{x}_p} (\partial_{\bar{x}} u)^2 d\bar{x}
=\int_0^{\bar{x}_p} (\partial_{\bar{x}} u)^2 d\bar{x}  .
\end{equation}
On the other hand, by integrating by parts
$$
F_{1,p+1} = \int_0^{\bar{x}_p} (\partial_{\bar{x}} \bar{x})u^{p+1} (\bar{x})d \bar{x} =  [\bar{x} u^{p+1} ]_0^{\bar{x}_p} - \int_0^{\bar{x}_p} \bar{x} \partial_{\bar{x}} 
(u^{p+1}) d\bar{x} = - (p+1)  \int_0^{\bar{x}_p} \bar{x} u^{p} \partial_{\bar{x}}u d\bar{x}  ,
$$
so that we get by
using again Emden equation (\ref{eq:emden}) and by integrating by parts:
$$
F_{1,p+1}  =  (p+1)  \int_0^{\bar{x}_p} \bar{x} \partial_{\bar{x}}^2 u \partial_{\bar{x}}u d\bar{x}
=
\frac{p+1}{2}\Big( \big[ \bar{x} (\partial_{\bar{x}} u)^2\big]_0^{\bar{x}_p} - \int_0^{\bar{x}_p} (\partial_{\bar{x}} u)^2 d \bar{x} \Big)  
=
\frac{p+1}{2}\Big( \bar{x}_p (\partial_{\bar{x}} u)^2({\bar{x}_p} )- \int_0^{\bar{x}_p} (\partial_{\bar{x}} u)^2 d \bar{x} \Big)  
.
$$
Substituting (\ref{eq:appD4}) into this equality gives:
\begin{equation}
\label{eq:app7}
F_{1,p+1} = \frac{p+1}{p+3} \bar{x}_p (\partial_{\bar{x}} u(\bar{x}_p))^2  .
\end{equation}
We have, by using Emden equation (\ref{eq:emden}) and by integrating by parts:
\begin{align*}
G_{1,p} &=  - \int_{-\bar{x}_p}^{\bar{x}_p} d\bar{x} u^p(\bar{x}) \Big( \int_{-\bar{x}_p}^{\bar{x}} d\bar{y}  (\bar{x}-\bar{y}) \partial_{\bar{y}}^2 u(\bar{y} )
+ \int_{\bar{x}}^{\bar{x}_p} d\bar{y}  (\bar{x}-\bar{y}) \partial_{\bar{y}}^2 u (\bar{y} )
\Big) \\
&=- \int_{-\bar{x}_p}^{\bar{x}_p} d\bar{x} u^p(\bar{x}) \Big(  \big[  (\bar{x}-\bar{y})  \partial_{\bar{y}}  u(\bar{y} ) \big]_{-\bar{x}_p}^{\bar{x}}
+ \int_{-\bar{x}_p}^{\bar{x}} d\bar{y} \partial_{\bar{y}} u(\bar{y} )
+ \big[  (\bar{y}-\bar{x}) \partial_{\bar{y}} u(\bar{y} ) \big]_{\bar{x}}^{\bar{x}_p}
- \int_{\bar{x}}^{\bar{x}_p} d\bar{y} \partial_{\bar{y}} u(\bar{y} )
\Big) \\
& = - 2 \int_{-\bar{x}_p}^{\bar{x}_p} d\bar{x} u^p(\bar{x}) \Big( \bar{x}_p \partial_{\bar{x}}u(\bar{x}_p) + u (\bar{x})\Big)  .
\end{align*}
Since $\int_{-\bar{x}_p}^{\bar{x}_p}  u^p(\bar{x}) d\bar{x}= - 2 \int_{0}^{\bar{x}_p}  \partial_{\bar{x}}^2 u(\bar{x}) d\bar{x} = - 2 \partial_{\bar{x}} u(\bar{x}_p)$ we get by (\ref{eq:app7}):
$$
G_{1,p} = 4 \Big( \frac{p+3}{p+1} -1 \Big) F_{1,p+1} = \frac{8}{p+1} F_{1,p+1},
$$
which is the desired result.

\subsection{Determination of the parameters of the stationary incoherent structure in the simulations}

The potential $V_L$ has mean zero so we have $V_L(x)=V_{st}(x) -  \left< V_{st}\right> $ and
\begin{align*}
{\cal H}_{nl,L} &=
 \frac{1}{2} \int_{[-L/2,L/2]}   \Big( V_{st}(x)  -  \left< V_{st}\right> \Big) N_{st}(x) dx , \quad \quad
 \left< V_{st}\right> = \frac{1}{L} \int_{[-L/2,L/2]} V_{st}(y)dy.
\end{align*}
The potential $V_{st}$ is equal to $- r_p^{-1/p} N_{st}(x)$ within the support $[-x_p,x_p]$ of $N_{st}$ ($x_p={\bar x}_p/\sqrt{\eta_p}$) and
it is a affine function outside of the support, which cancels at the boundary $x_p$ of the support 
and has constant derivative $\gamma {\cal M}$ (the derivative of $V_{st}$ is continuous and $\partial_x V_{st}(x_p) = 2\gamma \int_0^{x_p}N_{st}(x)dx = \gamma {\cal M}$).
We find 
\begin{align*}
\left< V_{st} \right>
&=
\frac{2}{L} \Big\{  \int_0^{{x}_p} V_{st}(x) dx +\int_{{x}_p}^{L/2} V_{st}(x) dx \Big\} \\
&=
\frac{2\gamma}{L}
\Big\{ - \frac{F_{1,1}}{4F_{1,p}^3} N_{st}(0)^{-2} {\cal M}^3 +\frac{1}{2} {\cal M}(\frac{L}{2}-{x}_p)^2\Big\}  .
\end{align*}
If we neglect the last term because ${x}_p$ is close to $L/2$, then we get
$$
{\cal H}_{nl,L} =
-\frac{\gamma}{4} \frac{F_{1,p+1}}{F_{1,p}^3} N_{st}(0)^{-1} {\cal M}^3 + \frac{\gamma}{4 L} \frac{F_{1,1}}{F_{1,p}^3}
N_{st}(0)^{-2} {\cal M}^4 .
$$
Using the value ${\cal H}_{nl,L}=- \gamma {\cal M}^2 L/24$ (by (\ref{eq:Hlnltheo})), this gives the relation
\begin{equation}
\label{eq:determin2}
-\frac{1}{6}= 
- \frac{F_{1,p+1}}{F_{1,p}^3}\Big(\frac{{\cal M}}{N_{st}(0)L}\Big)
 + \frac{F_{1,1}}{F_{1,p}^3}  \Big(\frac{{\cal M}}{N_{st}(0)L}\Big)^2.
\end{equation}
On the other hand, by (\ref{eq:calN_1d}) and (\ref{eq:calH_lin_1d}), 
$$
\frac{{\cal H}_{l}}{{\cal M}^3} = \frac{\gamma}{4} \frac{F_{1,p+1}}{(p+1) F_{1,p}^3} N_{st}(0)^{-1}  .
$$
Using the value ${\cal H}_{l}= \gamma {\cal M}^2 L/24$ (by (\ref{eq:Hlnltheo})) this gives the relation
\begin{equation}
\label{eq:determin3}
\frac{{\cal M}}{N_{st}(0)L}  = \frac{ (p+1)F_{1,p}^3 }{6F_{1,p+1}} .
\end{equation}
Substituting into (\ref{eq:determin2}) gives the algebraic equation
\begin{equation}
 \frac{6 F_{1,p+1}^2}{F_{1,1} F_{1,p}^3} = \frac{(p+1)^2}{6p} ,
\label{eq:alge}
\end{equation}
which makes it possible to determine $p$ and to obtain (\ref{eq:valuep}).
Then we get $N_{st}(0)$ from (\ref{eq:determin3}) which gives (\ref{eq:valueNst0}).
Finally note that using (\ref{eq:calN_1d}) and the expression of $r_p=\frac{d_0}{\sqrt{2\pi \alpha}} \frac{ \Gamma(p+1/2)}{\Gamma(p+1)}$, we obtain 
\begin{eqnarray}
\frac{N_{st}(0)}{d_0} = \Big( \frac{\gamma \bar{\rho} L^2F_{1,p}}{12 F_{1,p+1}}\Big)^p 
 \frac{(p+1)^p}{\sqrt{2\pi \alpha}} \frac{\Gamma(p+\frac{1}{2})}{\Gamma(p+1)}
 \simeq  8.6 \, 10^{-3}\,
 \frac{(\gamma \bar{\rho} L^2)^p }{\sqrt{\alpha}}. \nonumber
\end{eqnarray}

\section{The virial theorem for the WT-VPE in spatial dimension $d$}
\label{app:E}

The moment of inertia can be defined:
$$
I=\int |\bx|^2 N(\bx,t) d\bx= \frac{1}{(2\pi)^d} \iint |\bx|^2 n(\bk,\bx,t) d\bx d\bk .
$$
By taking time derivatives and using the WT-Vlasov Eq.(\ref{eq:vlasov_N0}
-\ref{eq:V_0_inc}), we get
$$
\partial_t^2 I = \frac{2\alpha^2}{(2\pi)^d} \iint  |\bk|^2 n(\bk,\bx,t) d\bx d\bk
-2\alpha \int \bx \cdot \partial_\bx V (\bx,t)N(\bx,t) d\bx  ,
$$
with $V(\bx,t) = - \gamma \int U_d(\bx-\by) N(\by,t) d\by$.
For definiteness, we recall that ${\cal H}_l(t) = \frac{1}{(2\pi)^d } \iint  \frac{\alpha}{2}|\bk|^2 n(\bk,\bx,t) d\bx d\bk$,  ${\cal H}_{nl}(t) = \frac{1}{2} \int V (\bx,t)N(\bx,t) d\bx$, and 
the mass ${\cal M} = \int N(\bx,t)  d\bx $. 

\subsection{The case $d=1$}
By taking into account $U_1(x)=-|x|$, we obtain
$$
 \int x \partial_x V (x,t)N(x,t) dx
=- \frac{\gamma}{2} \iint N(x,t) U_1(x-y) N(y,t)  dx dy.
$$
Therefore
$$
\frac{1}{2\alpha} \partial_t^2 I 
= 2{\cal H}_l(t)-{\cal H}_{nl}(t) .
$$
If the system is in a steady state, then $\partial_t^2 I=0$ and $ 2{\cal H}_l={\cal H}_{nl}$.

\subsection{The case $d=2$}
By taking into account $U_2(\bx)=-\log |\bx|$, we obtain 
\begin{align*}
 \int \bx \cdot \partial_\bx V (\bx,t)N(\bx,t) d\bx
&= \gamma \iint \bx \cdot \frac{\bx-\by}{|\bx-\by|^2}N(\by,t)  N(\bx,t)  d\bx d\by
= \frac{\gamma}{2} \iint (\bx -\by)\cdot \frac{\bx-\by}{|\bx-\by|^2}N(\by,t)  N(\bx,t)  d\bx d\by \\
&= \frac{\gamma}{2} {\cal M}^2
.
\end{align*}
Then 
$$
\frac{1}{4\alpha} \partial_t^2 I 
= {\cal H}_l(t)-\frac{\gamma}{4} {\cal M}^2  . 
$$
If the system is in a steady state, then $\partial_t^2 I=0$ and ${\cal H}_l= \frac{\gamma}{4} {\cal M}^2$.

\subsection{The case d=3}
By taking into account $U_3(\bx)=1/|\bx|$, and proceeding as for $d=2$, we obtain
$ \int \bx \cdot \partial_\bx V (\bx,t)N(\bx,t) d\bx = -{\cal H}_{nl}$.
Then
$$
\frac{1}{4\alpha} \partial_t^2 I = {\cal H}_l(t)+\frac{1}{2} {\cal H}_{nl} . 
$$
If the system is in a steady state, then $\partial_t^2 I=0$ and ${\cal H}_{nl}= -2{\cal H}_{l}$.

\section{The coherent soliton and incoherent structure for $d=3$}
\label{app:F}
\subsection{Dynamics of the coherent soliton ($d=3$)}
\label{sec:stat_sol_CS_3d}

For the sake of completeness, we recall in a simplified form the 3d single soliton solution of the effective SPE (\ref{eq:nls_A0}-\ref{eq:V_0_coh}) discussed in Ref.\cite{PRL21} in the case of binary solitons. 
The Lagrangian of the effective SPE (\ref{eq:nls_A0}-\ref{eq:V_0_coh}) is
$$
{\cal L} = \int  \frac{i}{2} \big( A \partial_t A^* -\partial_t A  A^*  \big)
+\frac{\alpha}{2}|\nabla A|^2 +\frac{1}{2} V_{S}(\bx) |A|^2 +\frac{2 \pi \gamma N({\bf 0})}{3} |\bx|^2 |A|^2  d\bx ,
$$
with 
$$
V_{S}(\bx) = - \gamma \int \frac{|A(\by)|^2}{|\bx-\by|} d\by .
$$
Following the variational  approach, we consider the Gaussian ansatz
\begin{eqnarray}
A(t,\bx) = a(t) \exp\Big( -\frac{|\bx -\bx_o(t)|^2}{2R^2(t)} +i b(t)|\bx -\bx_o(t)|^2 +i\boldsymbol{\kappa}(t)\cdot(\bx-\bx_o(t))+i \theta(t)\Big) .
\end{eqnarray}
We substitute the ansatz into the Lagrangian and calculate the effective Lagrangian in terms of
$a$, $R$, $b$, $\theta$, $\bx_o$, $\boldsymbol{\kappa}$ and  their  time  derivatives:
$$
{\cal L}  = \frac{3M_{S}R^2}{2} \partial_t b +M_{S} \big( \partial_t \theta
-\boldsymbol{\kappa}\cdot \partial_t \bx_o \big) +\frac{\alpha M_{S}}{4}
\big(\frac{3}{R^2}+12 b^2 R^2 + 2|\boldsymbol{\kappa}|^2 \big) - \frac{\gamma}{\sqrt{2\pi}}\frac{M_{S}^2}{R} + \pi \gamma N({\bf 0})  M_{S}
\big(R^2 + \frac{2}{3} |\bx_o|^2 \big),
$$  
where
$M_{S} = \int |A|^2 d\bx=\pi^{3/2} a^2 R^3$ is the conserved mass of the soliton.
The  evolution  equations  for  the
parameters of the ansatz are then derived from the effective
Lagrangian   by   using   the   corresponding   Euler-Lagrange
equations $\delta \int {\cal L} dt =0$. 
This first establishes a closed-form system of ordinary differential equations for the center $\bx_o$ and central wavevector $\boldsymbol{\kappa}$ of the soliton:
\begin{eqnarray}
\label{eq:center_3d}
&&\partial_t \bx_o =  \alpha \boldsymbol{\kappa} ,\\
&&\partial_t \boldsymbol{\kappa} = - \frac{4\pi}{3}  \gamma N({\bf 0}) \bx_o .
\label{eq:kappa_3d}
\end{eqnarray}
In particular, 
the motion of the soliton center $\bx_o$ is periodic (with period $\sqrt{3 \pi}/\sqrt{\alpha \gamma N({\bf 0})}$) and follows an ellipse.
This also yields  a  closed-form
ordinary differential equation for the width $R$:
\begin{equation}
\partial_t^2 R = \frac{\alpha^2}{R^3} - \frac{2 \alpha \gamma M_{S}}{3 \sqrt{2\pi} R^2} - \frac{4 \pi \gamma N({\bf 0})}{3} R .
\end{equation}
The energy is of the form
$ 
\frac{1}{2} (\partial_t R)^2 + W(R) ,
$
with the effective potential
$$
W(R) = \frac{\alpha^2}{2 R^2} - \frac{2 \alpha \gamma M_{S}}{3\sqrt{2\pi} R} +\frac{2 \pi \alpha\gamma N({\bf 0})}{3} R^2 .
$$
This equation has a stable equilibrium $R_S$ provided $\partial_R W(R_S)=0$ and $\partial_R^2 W(R_S)>0$.
For any positive mass $M_{S}$, this provides the unique stable solution with radius $R_S(M)$ that is the solution of Eq.(\ref{eq:masse_3d}), namely 
\begin{equation}
M_{S} = \frac{\sqrt{2\pi}}{\gamma}\Big( \frac{3\alpha}{2R_S}- 2 \pi \gamma N({\bf 0}) R_S^3\Big)  .
\label{eq:M_coh_3d}
\end{equation}

\subsection{Stationary solution for the incoherent structure ($d=3$)}
\label{sec:stat_sol_vlasov_3d}


The parabolic potential that confines the soliton in Eq.(\ref{eq:nls_A0}-\ref{eq:V_0_coh}) can be characterized by deriving a stationary solution of the  WT-VPE (\ref{eq:vlasov_N0}-\ref{eq:V_0_inc}).
Here we extend to $d=3$ the analysis reported in section~\ref{sec:stat_sol_vlasov_3d} for $d=1$.
Following the usual procedure \cite{binney}, we recall that any function $n_{st}(h)$ of the reduced hamiltonian $h=\frac{\alpha}{2}|\bk|^2+V_{st}(\bx)$ is a stationary solution of the WT-VPE (\ref{eq:vlasov_N0}-\ref{eq:V_0_inc}).
Assuming a spherically symmetric solution in $x=|\bx|$ and $k=|\bk|$, we have $N_{st}(x)=\frac{4\pi}{(2\pi)^3}\int_0^{k_c(x)} n_{st}(k,x) k^2 dk$ where the frequency cut-off is determined by the condition that the particles are trapped by the potential $h \le 0$, i.e., $k_c(x)=\sqrt{-2 V_{st}(x)/\alpha}$.
Then we have the self-consistent equations:
\begin{eqnarray}
N_{st}(x)&=& \frac{1}{\sqrt{2} \pi^2 \alpha^{\frac{3}{2}}} \int_{V_{st}}^0 n_{st}(h) \sqrt{h-V_{st}} dh,\\
(\partial_x^2 +2x^{-1}\partial_x)V_{st} &=& 4 \pi \gamma N_{st} ,
\end{eqnarray}
where $(\partial_x^2 +x^{-1}\partial_x)$ is the  spherically symmetric Laplacian.
We look for a solution in the form $n_{st}(h)=d_0 h^{p-3/2}$, where $d_0$ and $p$ are constant.
By integration we obtain 
\begin{eqnarray}
N_{st}(x)=r_p \big(-V_{st}(x)\big)^{p},
\end{eqnarray}
where 
$
r_p= \frac{d_0}{ (2 \pi \alpha)^{\frac{3}{2}}} \frac{\Gamma(p-\frac{1}{2})}{\Gamma(p+1)}.
$
Taking the Laplacian, we get the following form of the Emden equation for the IS $N_{st}(x)$:
\begin{eqnarray}
(\partial_x^2 +2x^{-1}\partial_x) \Big( N_{st}^{\frac{1}{p}} \Big)=-4\pi \gamma r_p^{\frac{1}{p}} N_{st}.
\label{eq:emden_N_0_3d}
\end{eqnarray}

We introduce the spatial variable ${\bar x}=\sqrt{\eta_p}x$ and amplitude ${u}({\bar x})=(N_{st}({\bar x}/\sqrt{\eta_p})/N_{st}(0))^{\frac{1}{p}}$, with $\eta_p=4 \pi \gamma r_p^{\frac{1}{p}} N_{st}^{1-\frac{1}{p}}(0)$ and $u({\bar x}) \ge 0$.
Eq.(\ref{eq:emden_N_0_3d}) then recovers the standard Emden equation
\begin{eqnarray}
(\partial_{\bar{x}}^2 + 2\bar{x}^{-1}\partial_{\bar{x}}) {u}({\bar x}) = - {u}({\bar x})^p  ,
\label{eq:emden_3d}
\end{eqnarray} 
with the initial conditions: ${u}(0)=1$ and $\partial_{\bar{x}} {u}(0)=0$.


We have a family of stationary solutions that depend on the three parameters $p$, $N_{st}(0)$ and $N_{st}(0)/d_0$.
Extending the analysis developed for $d=1$, we can compute their mass and energies.
The mass ${\cal M} = \int d\bx N_{st}(x) = 4\pi \int_0^\infty N_{st}(x) x^2 dx$ is 
\begin{eqnarray}
{\cal M} = N_{st}(0)^{\frac{3-p}{2p}}  (4\pi \gamma )^{-\frac{3}{2}}
\Big( \frac{d_0}{(2 \pi \alpha)^{\frac{3}{2}}}\frac{\Gamma(p-\frac{1}{2})}{\Gamma(p+1)}\Big)^{-\frac{3}{2p}}
F_{3,p},
\label{eq:masse_IS_3d}
\end{eqnarray} 
where $F_{3,p} = 4\pi \int_0^{{\bar x}_p} \bar{x}^2 u(\bar{x})^pd \bar{x}$ and $u(\bar{x})$ is the solution to Emden equation (\ref{eq:emden_3d}).
The total energy (Hamiltonian) of the WT-VPE (\ref{eq:vlasov_N0}-\ref{eq:V_0_inc}) has a linear contribution 
${\cal H}_{l} = 
\frac{1}{(2\pi)^3} \int d\bx \int d\bk   \frac{\alpha k^2}{2} n_{st}(k,x) =
\frac{(4\pi)^2 }{(2\pi)^3} \int_0^\infty x^2 dx \int_0^\infty k^2 dk   \frac{\alpha k^2}{2} n_{st}(k,x)$ and a nonlinear contribution 
${{\cal H}}_{nl} = - \frac{\gamma}{2} \iint d\bx d\by  \frac{N_{st}(x) N_{st}(y)}{|\bx-\by|} = -8\pi^2 \gamma
\int_0^\infty dx \int_0^\infty dy \, xy\min(x,y) N_{st}(x)N_{st}(y)$,
and we find
\begin{eqnarray}
\label{eq:Hl_IS_3d}
{\cal H}_l =  \frac{6 \pi}{p+1} N_{st}(0)^{\frac{5-p}{2p}} (4\pi \gamma )^{-\frac{3}{2}}
\Big( \frac{d_0}{(2 \pi \alpha)^{\frac{3}{2}}}\frac{\Gamma(p-\frac{1}{2})}{\Gamma(p+1)}\Big)^{-\frac{5}{2p}}
 F_{3,p+1} , \\
{\cal H}_{nl} =  -2\pi N_{st}(0)^{\frac{5-p}{2p}} (4\pi \gamma )^{-\frac{3}{2}}
\Big( \frac{d_0}{(2 \pi \alpha)^{\frac{3}{2}}}\frac{\Gamma(p-\frac{1}{2})}{\Gamma(p+1)}\Big)^{-\frac{5}{2p}}
  G_{3,p} ,
\label{eq:H_IS_3d}
\end{eqnarray}
where  $G_{3,p}=\int_0^{{\bar x}_p} d\bar{x} \int_0^{{\bar x}_p} d\bar{y} \,\bar{x}\bar{y}\min(\bar{x},\bar{y}) u(\bar{x})^p u (\bar{y})^p$
and $F_{3,p+1}=\int_0^{{\bar x}_p} d\bar{x}\, \bar{x}^2 u(\bar{x})^{p+1}$.
We have $\frac{6 F_{3,p+1}}{p+1}=G_{3,p}$, which shows that ${\cal H}_{nl}=-2 {\cal H}_{l}$. 
This is in agreement with the virial theorem.
Indeed, the virial theorem for the WT-VPE provides a general relation between the moment of inertia $I(t)=\int |\bx|^2 N(\bx,t) d\bx$, and the linear and nonlinear contributions to the Hamiltonian.
In a stationary state, $\partial_{t}^2 I=0$, the energies verify $2 {\cal H}_{l}+{{\cal H}}_{nl}=0$, see Appendix~\ref{app:E}.
This relation is indeed satisfied by the stationary solutions presented in this Appendix.

Finally, the radius $k_c=\sqrt{-2 V_{st}(0)/\alpha}$ of the spherically symmetric spectrum, which is compactly supported in $[0,k_c]$, is
\begin{eqnarray}
k_c = \sqrt{\frac{2}{\alpha}} N_{st}(0)^{\frac{1}{2p}}  
\Big( \frac{d_0}{(2 \pi \alpha)^{\frac{3}{2}}}\frac{\Gamma(p-\frac{1}{2})}{\Gamma(p+1)}\Big)^{-\frac{1}{2p}}.
\label{eq:kc_3d}
\end{eqnarray}

\subsection{Determination of the parameters of the stationary incoherent structure in the simulations}

As in the case $d=1$, the stationary solution is characterized by three parameters $p$, $N_{st}(0)$ and $N_{st}(0)/d$, and we have two independent equations (\ref{eq:masse_IS_3d}-\ref{eq:Hl_IS_3d}) in order to determine them from the initial conditions of the simulations. 
Indeed, the stationary state has known mass and kinetic energy, because ${\cal M}=\bar{\rho} L^3$ , ${\cal H}_l+{\cal H}_{nl}={\cal H} = -\gamma {\cal M}^2 c_{3,L}/2$ (since ${\cal H}_{l}(t=0)=0$ and ${\cal H}_{nl,L}(t=0)=0$)
and $2{\cal H}_l+{\cal H}_{nl}=0$ so that
\begin{equation}
{\cal M}=\bar{\rho}L^3, \quad \quad {\cal H}_l= \frac{2\pi c_3 \gamma {\cal M}^2}{L} = 2\pi c_3 \gamma \bar{\rho}^2 L^5 ,
\end{equation}
with $c_3 = \int_{[-1/2,1/2]^3} U_3(\bx) d\bx\simeq 3.17$.
Contrarily to the case $d=1$, we do not make the assumption that the stationary solution fills the box $[-L/2,L/2]^3$, because we do not have simulation results to support this hypothesis, and it seems very rough to assume that a spherically symmetric function fits a square box.
That means that the parameter $p$ cannot be determined from the initial conditions but from the observation
of the stationary solution.
Given the parameter $p$, the amplitude of the stationary solution can be determined from (\ref{eq:masse_IS_3d}-\ref{eq:Hl_IS_3d}) by eliminating $d_0$:
\begin{equation}
N_{st}(0)=\frac{{\cal M}}{L^3}  \frac{F_{3,p}^5 (p+1)^3 c_3^3}{(12 \pi F_{3,p+1})^3}.
\label{eq:N_0vsp_3d} 
\end{equation}

\end{widetext}




\begin{thebibliography}{99}
\bibitem{comment1}
The growth-rate of the gravitational instability in $\sigma_{\rm inst} = \sqrt{\alpha \gamma \eta_d {\bar \rho}-\alpha^2 k^4/4}$, so that modes with $k \ge k_J=2\pi/\Lambda_J$ with $\Lambda_J=\sqrt{2}\pi (\alpha/(\gamma \eta_d {\bar \rho}))^{1/4} \simeq \Lambda$ are stable.
\end{thebibliography}

\end{document}